\documentclass[aps,pra,superscriptaddress,singlecolumn,showkeys,notitlepage,nofootinbib,11pt,tightenlines,longbibliography]{revtex4-1}

\usepackage{appendix}

\makeatletter
\renewcommand{\p@subsection}{}
\renewcommand{\p@subsubsection}{}
\makeatother

\usepackage{appendix}
\usepackage[table]{xcolor}
\newcommand{\nocontentsline}[3]{}

\usepackage{graphicx, color, graphpap}% Include figure files
\usepackage{enumitem}
\usepackage{amssymb}
\usepackage{amsthm}
\usepackage[OT1]{fontenc} 
\usepackage{multirow}
\usepackage[margin=0.9in]{geometry}
\usepackage[colorlinks=true,citecolor=blue,linkcolor=blue]{hyperref}
\usepackage[T1]{fontenc}

\usepackage{bbm}
\usepackage{thmtools,thm-restate}
\usepackage{verbatim}
\usepackage{mathtools}
\usepackage{titlesec}
\usepackage{amsmath}
\usepackage{tikz}
\usepackage{soul}
\usetikzlibrary{quantikz}
\usepackage[caption = false]{subfig}
\usepackage{float}

\usepackage{tabularx}
\usepackage{lmodern,adjustbox}
\usepackage[skins,breakable]{tcolorbox}
\tcbuselibrary{listings}
\tcbuselibrary{breakable}
\newtcolorbox{Code}{enhanced,fonttitle=\sffamily\bfseries\large,valign=center
, drop fuzzy shadow,sidebyside,lefthand ratio=0.4,lower separated=false}

\long\def\ca#1\cb{} %Use for commenting out: \ca...\cb

\renewcommand{\geq}{\geqslant}
\renewcommand{\leq}{\leqslant}

\newcolumntype{s}{>{\columncolor[HTML]{AAACED}} p{3cm}}

\renewcommand{\vec}[1]{\boldsymbol{#1}}  % Bold vectors instead of arrow vectors

\newcommand{\thv}{\vec{\theta}}

{}%         Body font
{}%         Indent amount (empty = no indent, \parindent
\begin{document}

\title{GPT on a Quantum Computer}

\author{Yidong Liao}
\email{yidong.liao@student.uts.edu.au}
\affiliation{Centre for Quantum Software and Information, University of Technology Sydney, Sydney, NSW, Australia}
\affiliation{Sydney Quantum Academy, Sydney, NSW, Australia}
\author{Chris Ferrie} \email{christopher.ferrie@uts.edu.au}
\affiliation{Centre for Quantum Software and Information, University of Technology Sydney, Sydney, NSW, Australia}
\date{March 14, 2024}
\begin{abstract}
Large Language Models (LLMs) such as ChatGPT have transformed how we interact with and understand the capabilities of Artificial Intelligence (AI).  However, the intersection of LLMs with the burgeoning field of Quantum Machine Learning (QML) is only in its nascent stages. This paper presents an exploration of this niche by detailing a comprehensive framework for implementing the foundational Transformer architecture --- integral to ChatGPT --- within a quantum computing paradigm. We meticulously design quantum circuits that implement adapted versions of the transformer's core components and the generative pre-training phase.  By integrating quantum computing with LLMs, we aspire to open new avenues for research in QML and contribute to the ongoing evolution of AI technologies.
\end{abstract}
\maketitle

\begin{figure}[h!]
    \centering
\includegraphics[width=0.79\linewidth]{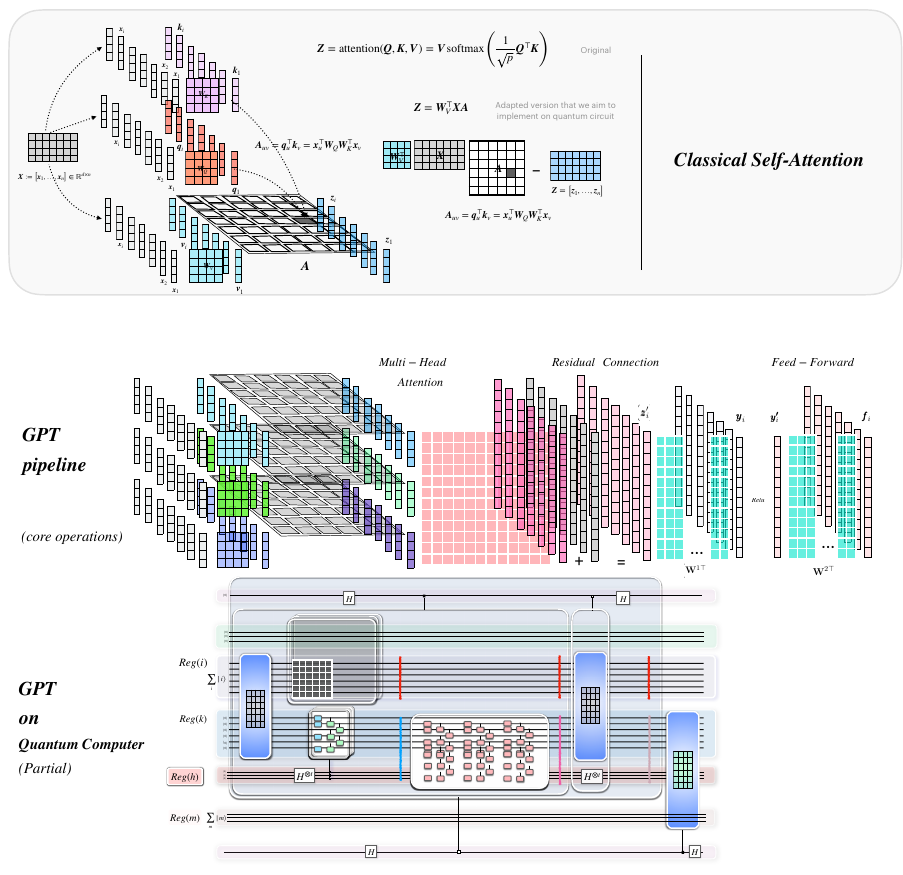}
  \caption{\textit{Preview of some key contents in this paper}. Graphical representations of the classical Self-Attention Mechanism, GPT pipeline, and overall quantum implementation of GPT. }
\end{figure}

\tableofcontents

\section{Introduction}\label{intro}
The emergence and rapid advancement of Large Language Models (LLMs) such as ChatGPT has had a significant global impact, revolutionizing our interactions with artificial intelligence and expanding our understanding of its capabilities. Models like GPT-4 have demonstrated the vast potential of LLMs in a wide range of applications across various domains. In the field of natural language processing (NLP), LLMs have proven to be highly proficient in tasks such as machine translation, sentiment analysis, question answering, and text summarization. They excel in identifying intricate language patterns, comprehending context, and generating text that is coherent and contextually appropriate.\newline

Concurrently, the field of quantum computing has seen remarkable progress, offering unprecedented computational power that promises to solve complex problems beyond the reach of classical computing~\cite{scholten2024assessing}. Quantum Machine Learning (QML), an intersection of quantum computing and machine learning, has emerged as a fertile ground for research, aiming to leverage quantum algorithms to enhance machine learning tasks~\cite{cerezo2022challenges}. Despite the significant achievements in QML, integrating quantum computing with state-of-the-art machine learning models, especially LLMs, is only at its nascent stages. Recent explorations such as Ref.~\cite{liu2024towards,gao2023fast} indicate a burgeoning interest in leveraging quantum computing to elevate the capabilities of LLMs.\newline

%Recent studies have begun to pave the way, demonstrating innovative methodologies for the integration of quantum computing principles with state-of-the-art machine learning models. Our research is positioned within this emerging landscape, seeking to bridge the gap by specifically focusing on the Generative Pre-trained Transformer (GPT) — the foundational version of ChatGPT. This paper presents a comprehensive framework for adapting key components of the GPT architecture for quantum computing, marking a significant step towards harnessing quantum computing's capabilities to enhance LLMs.This gap signifies a potential for an important area of research that could unveil new dimensions of computational efficiency and model performance. \newline

 This paper delves into the quantum implementation of Generative Pre-trained Transformer (GPT) \cite{radford2018improving} (also referred to as GPT-1\footnote{In this paper, we use "GPT" instead of "GPT-1."}) --- the original version of ChatGPT, focusing on adapting its architecture and pre-training processes to leverage the computational paradigms of quantum computing. We explore how quantum algorithms can be applied to the foundational components of GPT's architecture and the generative pre-training, with the potential to enhance its efficiency and capabilities. By integrating quantum computing with LLMs, we aspire to open new avenues for research in QML and contribute to the ongoing evolution of AI technologies. \newline

The rest of this paper is organized as follows: Section \ref{Background} provides a background on the transformer architecture used in GPT and the basics of language modeling. Section \ref{implement} details our approach to implement GPT's architecture on quantum computers, including the quantum circuit designs for key model components and the generative pre-training. Section \ref{conclusion} concludes the paper with a summary of our findings and directions for future research.

\section{Background}\label{Background}
\subsection{Transformer Architecture Used in GPT}\label{tran}

The Generative Pre-trained Transformer (GPT) \cite{radford2018improving} is the inaugural version in the series of groundbreaking language models developed by OpenAI, marking the beginning of a new era in NLP. GPT's architecture is predicated on the transformer model~\cite{vaswani2017attention}, a type of neural network that relies on self-attention mechanisms to process sequences of data, such as text. With 117 million parameters, GPT was a large model for its time, capable of capturing complex language patterns and generating coherent and contextually relevant text. GPT's introduction was a pivotal moment in NLP; it paved the way for the development of more advanced models, such as GPT-2 and GPT-3, setting the stage for the rapid advancement of AI technologies in the years that followed. The remainder of this section gives an overview of GPT's architecture and its training, and the next section \ref{llm} presents its application in language modeling.\newline

\begin{figure}[h!]
    \centering
    \includegraphics[width=0.96\linewidth]{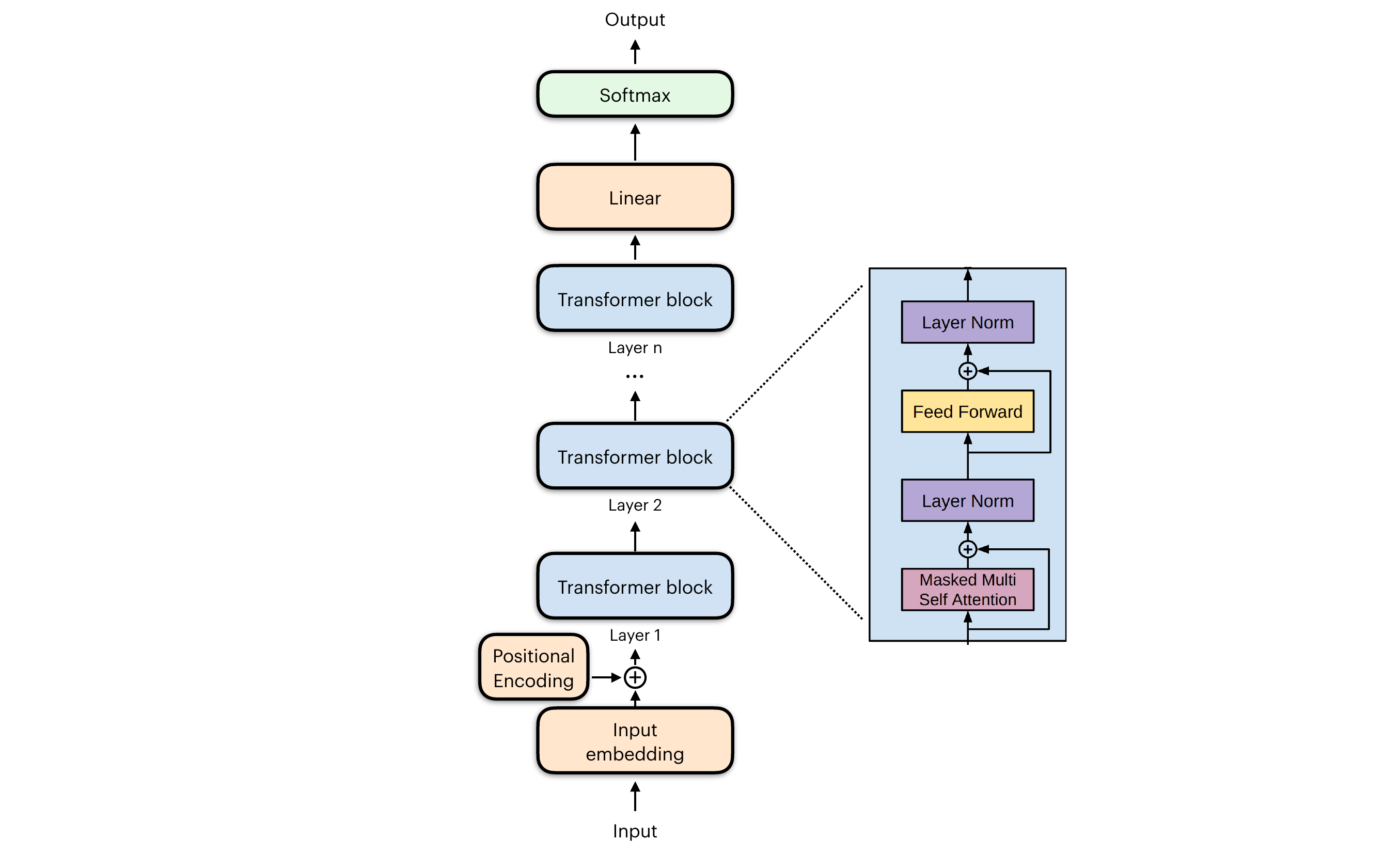}
   \caption{\textit{GPT's Architecture, adapted from \cite{radford2018improving}}. GPT's architecture is a multi-layer decoder-only Transformer \cite{liu2018generating} (a variant of the Transformer\cite{vaswani2017attention}). The primary part of the architecture is a stack of transformer blocks \cite{radford2018improving}, each of which is composed of two main components: a (masked) multi-head self-attention mechanism followed by a position-wise fully connected feed-forward network. Layer Normalization and Residual Connections are placed around these two main components.  The transformer blocks are stacked on top of each other, with each layer processing the output of the previous one. Prior to the input embedding entering the transformer blocks, positional encoding is added. 
}
    \label{gpt1}
\end{figure}

GPT's architecture is a multi-layer decoder-only Transformer \cite{liu2018generating} (a variant of the Transformer\cite{vaswani2017attention}). The primary part of the architecture is a stack of transformer blocks \cite{radford2018improving}, each of which is composed of two main components: a (masked) multi-head self-attention mechanism followed by a position-wise fully connected feed-forward network. Layer Normalization and Residual Connections are placed around these two main components.  The transformer blocks are stacked on top of each other, with each layer processing the output of the previous one. Prior to the input embedding entering the transformer blocks, positional encoding is added. Fig.\ref{gpt1} illustrates these components in GPT's architecture, the following paragraphs briefly\footnote{Here in this section, for each component in GPT's architecture, we only give an overview, the detailed mathematical descriptions of each component are presented in Section \ref{implement}.} introduce each component.  \newline

\paragraph{Multi-Head Masked Self-Attention Mechanism}. The multi-head self-attention mechanism, with the addition of a masking operation, is a core component of the transformer block in GPT. This attention mechanism allows a model to weigh the importance of different words in a sentence. Unlike previous models (such as Recurrent Neural Networks(RNNs) \cite{medsker1999recurrent}) that process words in a sequential manner, self-attention enables the model to look at all parts of the sentence simultaneously. This allows for a more nuanced understanding of context and relationships between words, regardless of their position in the sentence. The masking operation is a critical aspect of this mechanism, especially in the context of language modeling(a brief introduction is given in Section \ref{llm}): it ensures that the prediction of a current word does not get influenced by future words. \cite{ghojogh2020attention}\newline

\paragraph{Layer Normalization and Residual Connections}.
Each transformer block in GPT includes layer normalization and residual connections. Layer Normalization is applied after the self-attention mechanism and after the feed-forward network within each transformer block. It normalizes the inputs across the features, improving the stability of the model.
Residual Connections allow the input of each sub-layer (i.e., the self-attention and feed-forward networks) to be added to its output. \newline

\paragraph{Position-Wise Fully Connected Feed-Forward Network}.  In each transformer block in GPT, after the attention mechanism together with corresponding Layer Normalization and Residual Connection, the output is passed through a feed-forward network that applies the same transformation to each position separately and identically. \newline

\paragraph{Positional Encoding}.
Since GPT (and transformer models in general) does not inherently process sequential data in order, it uses positional encodings to incorporate information about the order of the sequence into its inputs. These positional encodings are added to the input embeddings at the bottom of the model stack, providing the model with information about the position of each word in the sequence. \newline

The training process of GPT consists of two main stages \cite{radford2018improving}: unsupervised pre-training and supervised fine-tuning. During pre-training, GPT is exposed to a large corpus of text data, learning the underlying structure of the language without any task-specific instructions. This stage allowed the model to develop a broad understanding of language, including grammar, semantics, and common phrases. The fine-tuning stage then adapted the pre-trained model to specific tasks, such as translation, question-answering, and summarization, by training it on smaller, task-specific datasets.

\subsection{Language Modeling basics}\label{llm}

This section briefly introduces one of GPT's applications in language modeling --- text generation, which is the foundation of the services provided by ChatGPT. We start by presenting an overview of language modelling:\newline

Language modeling is a fundamental aspect of natural language processing (NLP) that focuses on the development of probabilistic models capable of understanding, generating, and interpreting human language. At its core, a language model predicts the likelihood of upcoming sequences of words occurring in a text~\cite{JurafskyMartin}. This predictive capability enables a wide range of applications, from autocomplete systems in smartphones and email platforms to sophisticated chatbots, machine translation, speech recognition, and even content generation tools.\newline

The essence of language modeling lies in capturing the statistical structure of language—learning how words and phrases tend to come together to convey meaning. Early language models were relatively simple $n$-gram models\cite{brown1992class}, where the prediction of the next word in a sequence depended on the previous $n-1$ words. However, these models had limitations, particularly in dealing with long-term dependencies and the vast diversity of linguistic contexts. The advent of neural networks brought a significant leap forward in language modeling. Recurrent Neural Networks (RNNs)\cite{medsker1999recurrent}, and later, Long Short-Term Memory (LSTM) networks\cite{hochreiter1997long}, provided mechanisms to remember information over longer sequences, improving the model's ability to handle context in language. Despite these advances, RNNs and LSTMs also faced challenges, such as difficulty in training over very long sequences and capturing complex dependencies.\newline

The breakthrough came with the introduction of the Transformer architecture\cite{vaswani2017attention}, which led to the development of models like OpenAI's GPT series that demonstrated unprecedented capabilities in generating coherent and contextually relevant text over extended passages. The development of language models continues to be a vibrant area of research in AI, with ongoing work aimed at improving their accuracy, efficiency, and ability to understand and generate human language in all its complexity.\newline

 Next, we delve into fundamental concepts of language modeling, such as tokenization, explaining how a language model (here, GPT) operates for text generation.

\subsubsection{Tokenization, Word Embedding}

Tokenization is the process of converting text into tokens which are the basic units of language models. There are three levels of tokenization:

\begin{itemize}
  \item Character-level: Processes text one letter at a time.
  \item Word-level: Segments text into individual words.
  \item Subword-level: Breaks down words into subunits. For example, the subword tokenization of the phrase "Language Model" may look like ["Lan", "gu", "age ", "Mod", "el"].
\end{itemize}

Upon establishing the tokens for a language model, we arrange them into a structured vocabulary, assigning a distinct index to each token. These indices are then transformed into input features through various methodologies. Directly inputting these indices into the model is inadvisable, as the sequential order within the vocabulary does not inherently reflect semantic relationships. An alternative is the utilization of one-hot encoding. For a vocabulary encompassing 10,000 words, each word is symbolized by a 10,000-element vector, predominantly composed of zeros, save for the element corresponding to the word's indexed position. The primary benefit of one-hot encoding is its ability to preclude presumptions about word importance, facilitating the model's learning of word relationships during its training phase.\newline

\begin{figure}[h!]
    \centering
    \includegraphics[width=0.9\linewidth]{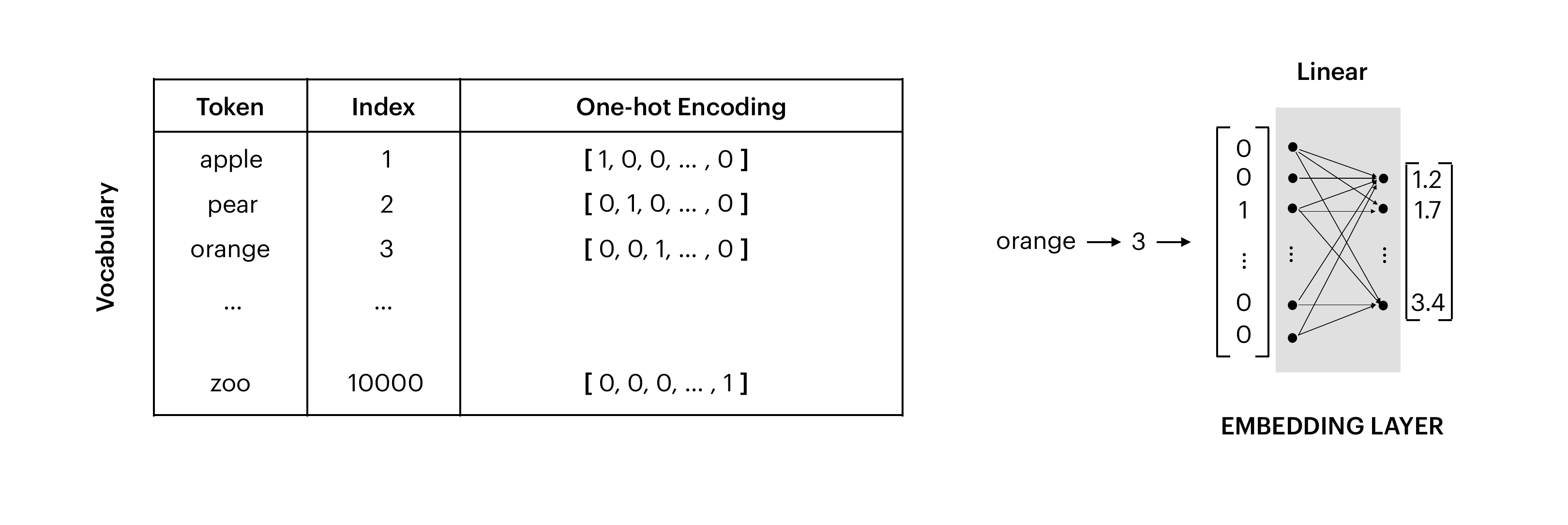}
   \caption{\textit{Tokenization, one-hot encoding and word embedding}.  }
    \label{onehot}
\end{figure}

However, one-hot encoding presents scalability issues in the context of extensive vocabularies. Considering the English language, with its repertoire exceeding 100,000 words, representing a single word would necessitate a vector comprising 100,000 elements. In scenarios involving lengthy sequences, this approach demands substantial storage space and computational resources. To mitigate this issue of dimensionality, the use of word embeddings is proposed\cite{mikolov2013efficient,pennington2014glove,goldberg2014word2vec,mikolov2015computing}. In this framework, an embedding layer projects the one-hot encoded tokens into a more condensed vector space. These embeddings, essentially denser token representations, are generated through a linear layer equipped with a weight matrix, which the model optimizes during its training process.  Fig.\ref{onehot} summarizes this section.

\subsubsection{Next word prediction}

\begin{figure}[h!]
    \centering
    \includegraphics[width=\linewidth]{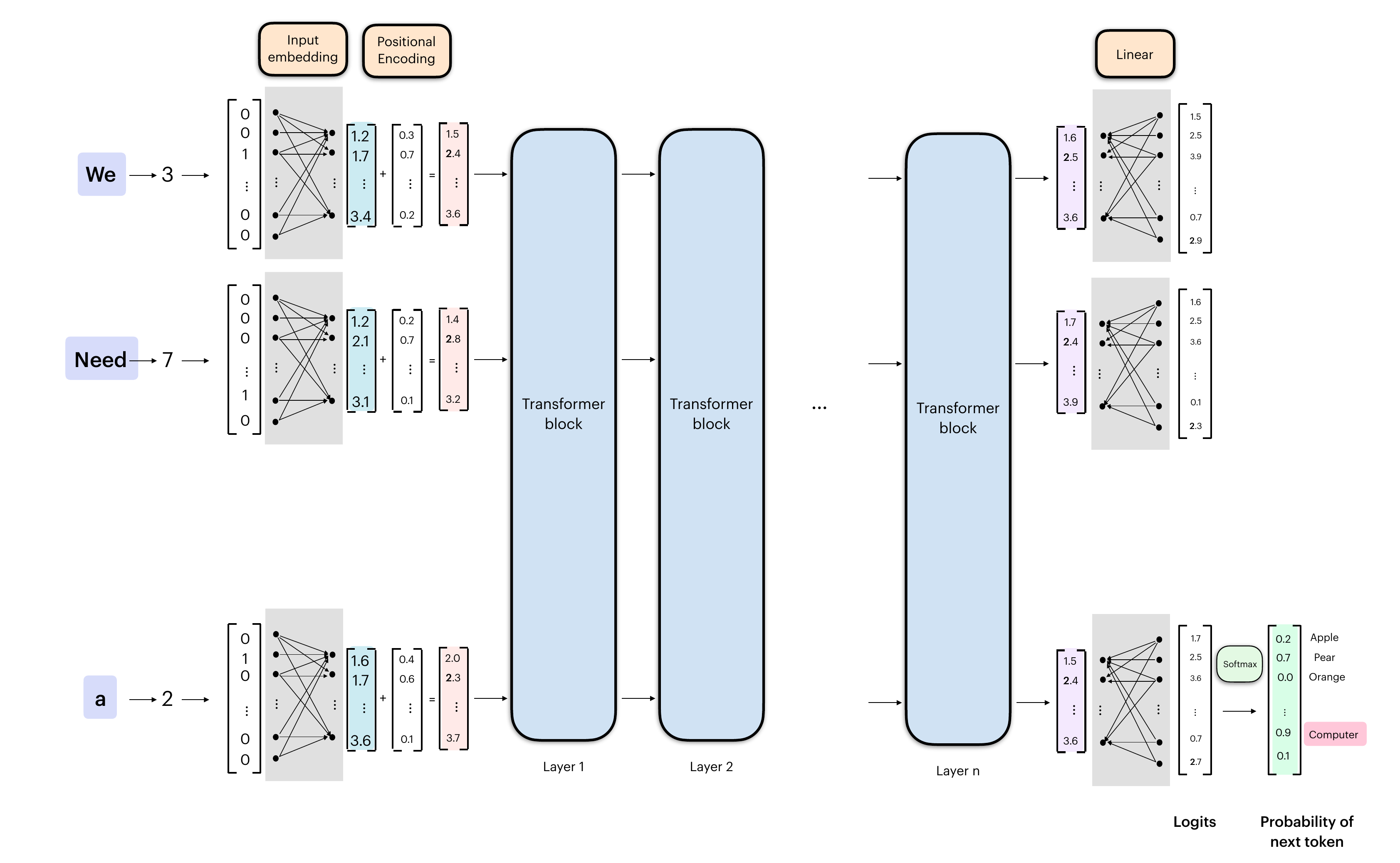}
   \caption{\textit{Next word prediction process of GPT}. In the inference stage, a language model (here, GPT) takes in a sequence of one-hot encoded tokens, and generates predictions for the next word in a sequence. The sequence of one-hot encoded tokens is first transformed through word embedding, as described in the above section. These embeddings, after the positional encodings are added, are then input into the transformer blocks. Then, a final linear layer is applied to map the outputs from the transformer blocks back into the vocabulary space, generating a sequence of transformed vectors. The last transformed vector in the sequence is referred as "logits". The logits are passed through a softmax activation function, yielding a probability distribution across the vocabulary, indicating the likelihood of each word as the next sequence component.
 }
    \label{llm12}
\end{figure}

In the inference stage, a language model (here, GPT) takes in a sequence of one-hot encoded tokens, and generates predictions for the next word in a sequence.\footnote{Here, we only consider autoregressive language models.} The sequence of one-hot encoded tokens is first transformed through word embedding, as described in the above section. These embeddings, after the positional encodings are added, are then input into the transformer blocks. Then, a final linear layer is applied to map the outputs from the transformer blocks back into the vocabulary space, generating a sequence of transformed vectors. The last transformed vector in the sequence is referred as "logits". The logits are passed through a softmax activation function, yielding a probability distribution across the vocabulary, indicating the likelihood of each word as the next sequence component. Fig.\ref{llm12} summarizes this section.

\subsubsection{Generative Pre-training}

GPT model undergo extensive pre-training on large text corpora using the following loss function:

\begin{equation}
    L(\theta) = -\sum_{t} \log P(w_t | w_{1:t-1}; \theta)
\end{equation}
where \(w_t\) is the \(t\)-th word, and \(\theta\) represents the model parameters. This pre-training endows GPT with a broad understanding of language, which is then refined for specific tasks through fine-tuning.\newline

 Given the unlabeled nature of these sentences, this process is classified as unsupervised learning. It involves the pairing of a text segment as input with its subsequent segment as the target. The training process encompasses processing these input-target pairs in batches. The loss is computed by evaluating the next token in the target output, and this process is repeated for each subsequent token in the sequence. The cumulative loss is calculated across all batches, followed by the execution of backpropagation to adjust the model's parameters.\newline

The culmination of this process is a pre-trained language model, which we can then employ for text generation. This begins with the input of an initial word or phrase, serving as the genesis for text generation. The model assesses this input to predict the next token, which is subsequently reintroduced into the model as the new input. This iterative process engenders a feedback loop, enabling the model to generate continuous text sequences.

\section{GPT on a Quantum Computer}\label{implement}

As outlined in Section \ref{tran} and depicted in Figure \ref{gpt1}, the architecture of GPT encompasses several key elements: input embedding, positional encoding, a series of transformer blocks, and a concluding linear layer followed by a softmax function. Within the context of this paper, it is assumed that both the input embedding and positional encoding are executed on classical computers. Our focus, however, shifts to detailing the implementation of the transformer blocks' core components on a quantum computer. Additionally, we delve into the methodologies employed for executing Generative Pre-training of the model on a quantum computer. To ensure a holistic presentation, a detailed exposition on the quantum implementation of positional encoding is provided in Appendix \ref{positional}.

\subsection{Input encoding}\label{input}
In this section, we focus on the process of input encoding for the quantum implementation of a  Transformer block. As illustrated in Figure \ref{llm12}, the input to the Transformer block is a sequence of vectors $\left\{\boldsymbol{x}_{i} \in \mathbb{R}^{d}\right\}_{i=1}^{n}$ stacked as a matrix $\boldsymbol{X}:=\left[\boldsymbol{x}_{1}, \ldots, \boldsymbol{x}_{n}\right] \in \mathbb{R}^{d \times n}$. It can be encoded in a quantum state (after normalization\footnote{Note here and throughout the paper, we omit the normalization factors.}) $\left|\psi_{\boldsymbol{X}}\right\rangle$ as,

\begin{equation}
\left|\psi_{\boldsymbol{X}}\right\rangle:=\sum_{i=1}^{n} |i\rangle\ket{\boldsymbol{x}_{i}}
\label{psix},
\end{equation}
where $\ket{\boldsymbol{x}_{i}}:=\sum_{k=1}^{d}\boldsymbol{x}^{(k)}_{i}|k\rangle$ is the amplitude encoding of the vector $\boldsymbol{x}_{i}$ whose $k$-th elements are denoted as $\boldsymbol{x}^{(k)}_{i}$. \newline

The entire state is prepared on two quantum registers hosting the index $k$ and index $i$, which are denoted as $Reg(k)$ and $Reg(i)$, respectively. The unitary that realizes the data encoding as,
\begin{equation}
U_{\boldsymbol{X}}: |i\rangle \ket{0} \to |i\rangle\ket{\boldsymbol{x}_{i}}, \forall i=1,\cdots n,
\label{Ux3}
\end{equation}
is represented as the blue box in Fig.\ref{qattention1}. $U_{\boldsymbol{X}}$ can be achieved by ``Controlled Quantum State Preparation(CQSP)'' process \cite{yuan2023optimal}.

\subsection{Attentions on Quantum Computer}

As mentioned in Section.~\ref{tran}, the multi-head self-attention, with the addition of a masking operation, is a core component of the transformer block in GPT. In this section, we present the quantum implementation of an adapted version of this component: Subsection.~\ref{selfsection} illustrates the adaptation of single-head self-attention and its implementation on quantum computers, while Subsection.~\ref{sectionmulti} discusses multi-head self-attention and its implementation on quantum computers. The masking operation and its quantum implementation are given in Appendix.~\ref{mask}.\newline

 In classical transformer architecture,  the attention function can be described as mapping queries, keys, and values to an output, where the queries, keys, values, and output are all vectors \cite{vaswani2017attention}. The query $\boldsymbol{q}_{i}$, key $\boldsymbol{k}_{i}$, and value $\boldsymbol{v}_{i}$ are $p$-dimensional, $p$-dimensional, and $r$-dimensional vectors defined as \cite{ghojogh2020attention}:

\begin{align}
& \mathbb{R}^{p} \ni \boldsymbol{q}_{i}=\boldsymbol{W}_{Q}^{\top} \boldsymbol{x}_{i}, \label{values}\\
& \mathbb{R}^{p} \ni \boldsymbol{k}_{i}=\boldsymbol{W}_{K}^{\top} \boldsymbol{x}_{i}, \label{key}\\
& \mathbb{R}^{r} \ni \boldsymbol{v}_{i}=\boldsymbol{W}_{V}^{\top} \boldsymbol{x}_{i},\label{quary}
\end{align}

where $\boldsymbol{W}_{Q} \in \mathbb{R}^{d \times p}, \boldsymbol{W}_{K} \in \mathbb{R}^{d \times p}$, and $\boldsymbol{W}_{V} \in$ $\mathbb{R}^{d \times r}$ are trainable matrices.\footnote{The "transpose" in \ref{quary},\ref{key},\ref{values} are in the definition for some reason which will be clear in Section \ref{ffn}.} Similar to vectors $\left\{\boldsymbol{x}_{i} \in \mathbb{R}^{d}\right\}_{i=1}^{n}$ being stacked as a matrix $\boldsymbol{X}:=\left[\boldsymbol{x}_{1}, \ldots, \boldsymbol{x}_{n}\right] \in \mathbb{R}^{d \times n}$, we define $\boldsymbol{Q}:=\left[\boldsymbol{q}_{1}, \ldots, \boldsymbol{q}_{n}\right] \in \mathbb{R}^{p \times n}$, $\boldsymbol{K}:=\left[\boldsymbol{k}_{1}, \ldots, \boldsymbol{k}_{n}\right] \in \mathbb{R}^{p \times n}$, and $\boldsymbol{V}:=\left[\boldsymbol{v}_{1}, \ldots, \boldsymbol{v}_{n}\right] \in$ $\mathbb{R}^{r \times n}$. From these definitions, we have

\begin{align}
& \boldsymbol{Q}=\boldsymbol{W}_{Q}^{\top} \boldsymbol{X}, \label{values2}\\
& \boldsymbol{K}=\boldsymbol{W}_{K}^{\top} \boldsymbol{X}, \label{key2}\\
& \boldsymbol{V}=\boldsymbol{W}_{V}^{\top} \boldsymbol{X},\label{quary2}
\end{align}

The "Scaled Dot-Product Attention" defined in \cite{vaswani2017attention} can be written in matrix form as:

\begin{equation}
	 \boldsymbol{Z}_0  :=\operatorname{attention}(\boldsymbol{Q}, \boldsymbol{K}, \boldsymbol{V})  =\boldsymbol{V} \operatorname{softmax}\left(\frac{1}{\sqrt{p}} \boldsymbol{Q}^{\top} \boldsymbol{K}\right),
\label{attentiondef}
\end{equation}

where $\boldsymbol{Z}_0=\left[\boldsymbol{z_0}_{1}, \ldots, \boldsymbol{z_0}_{n}\right]\in \mathbb{R}^{r \times n} $ .\newline

 Note that for each $i$, the queries, keys, and values are all from the same vector $\boldsymbol{x}_{i}$ in the sequence; this type of attention is referred to as the "self-attention" \cite{ghojogh2020attention}.
\begin{figure}[h!]
    \centering
    \includegraphics[width=\linewidth]{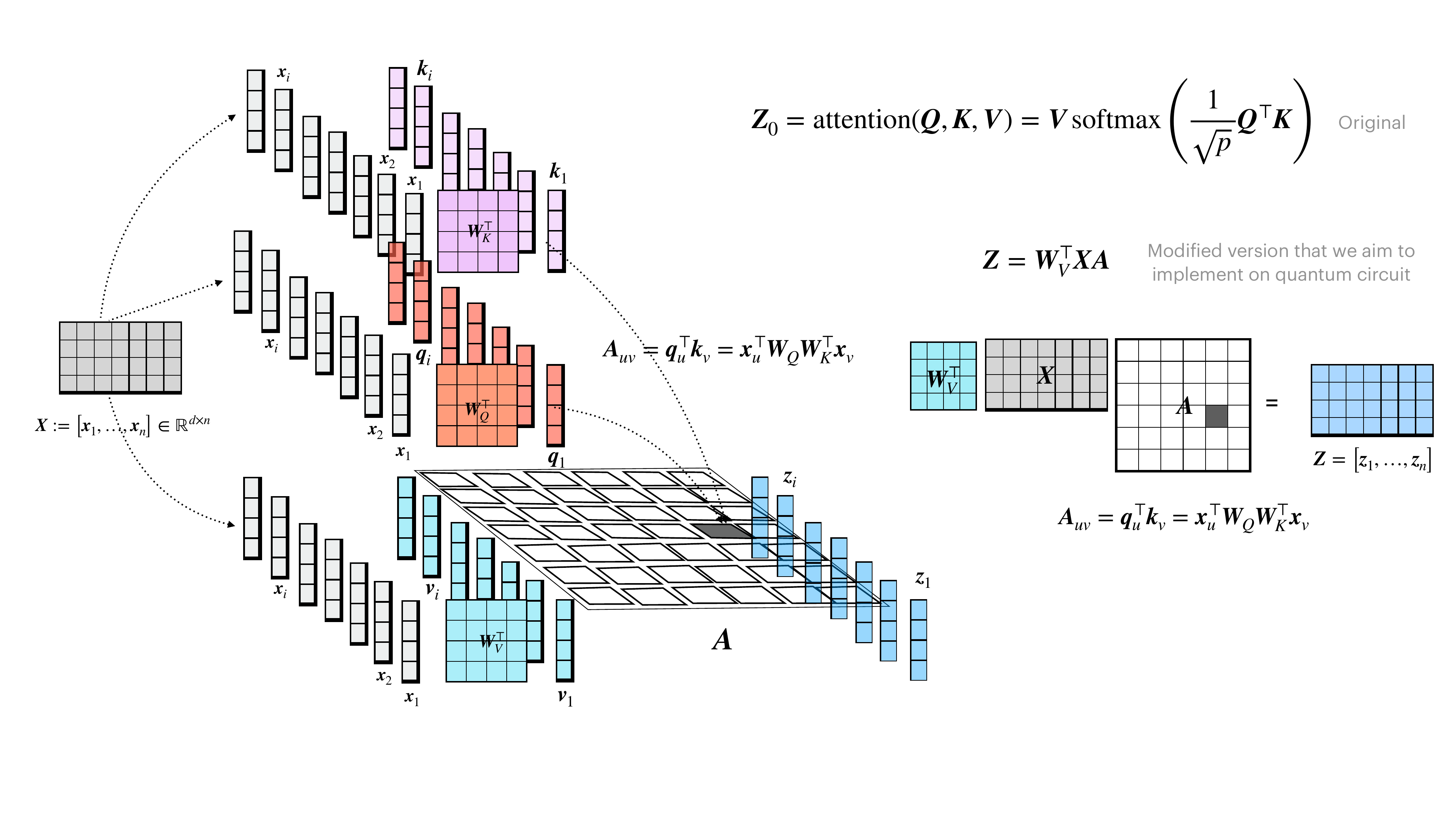}
   \caption{\textit{The Classical Self-Attention(modified version that we aim to implement on the quantum circuit)} As described in Ref.~\cite{vaswani2017attention,ghojogh2020attention}, an attention function maps queries, keys, and values to an output, with all being vectors. The queries $\boldsymbol{q}_{i}$, keys $\boldsymbol{k}_{i}$, and values $\boldsymbol{v}_{i}$ are $p$-dimensional, $p$-dimensional, and $r$-dimensional vectors, respectively, defined as: $\mathbb{R}^{p} \ni \boldsymbol{q}_{i}=\boldsymbol{W}_{Q}^{\top} \boldsymbol{x}_{i}, \mathbb{R}^{p} \ni \boldsymbol{k}_{i}=\boldsymbol{W}_{K}^{\top} \boldsymbol{x}_{i}, \mathbb{R}^{r} \ni \boldsymbol{v}_{i}=\boldsymbol{W}_{V}^{\top} \boldsymbol{x}_{i}$, where $\boldsymbol{W}_{Q} \in \mathbb{R}^{d \times p}, \boldsymbol{W}_{K} \in \mathbb{R}^{d \times p}$, and $\boldsymbol{W}_{V} \in \mathbb{R}^{d \times r}$ are the projection matrices. Similar to vectors $\{\boldsymbol{x}_{i} \in \mathbb{R}^{d}\}_{i=1}^{n}$ being stacked as a matrix $\boldsymbol{X}:=\left[\boldsymbol{x}_{1}, \ldots, \boldsymbol{x}_{n}\right] \in \mathbb{R}^{d \times n}$, we define $\boldsymbol{Q}:=\left[\boldsymbol{q}_{1}, \ldots, \boldsymbol{q}_{n}\right] \in \mathbb{R}^{p \times n}$, $\boldsymbol{K}:=\left[\boldsymbol{k}_{1}, \ldots, \boldsymbol{k}_{n}\right] \in \mathbb{R}^{p \times n}$, and $\boldsymbol{V}:=\left[\boldsymbol{v}_{1}, \ldots, \boldsymbol{v}_{n}\right] \in \mathbb{R}^{r \times n}$, respectively. The "Scaled Dot-Product Attention" defined in \cite{vaswani2017attention} can be written in matrix form as: $\boldsymbol{Z}_0 :=\operatorname{attention}(\boldsymbol{Q}, \boldsymbol{K}, \boldsymbol{V}) =\boldsymbol{V} \operatorname{softmax}\left(\frac{1}{\sqrt{p}} \boldsymbol{Q}^{\top} \boldsymbol{K}\right)$. Notably, for each $i$, the queries, keys, and values are derived from the same vector $\boldsymbol{x}_{i}$, a process referred to as "self-attention" \cite{ghojogh2020attention}. Denote $\operatorname{softmax}\left(\frac{1}{\sqrt{p}} \boldsymbol{Q}^{\top} \boldsymbol{K}\right)$ as $\boldsymbol{A}_0$ and substituting $\boldsymbol{V}=\boldsymbol{W}_{V}^{\top} \boldsymbol{X}$, we obtain $\boldsymbol{Z}_0 =\boldsymbol{W}_{V}^{\top} \boldsymbol{X} \boldsymbol{A}_0$. Considering the implementation of the $\operatorname{softmax}$ function via a quantum circuit is not straightforward and that scaling is managed in the block-encoding process, we aim to implement an alternative version of $\boldsymbol{A}_0$, denoted as $\boldsymbol{A}=\boldsymbol{Q}^{\top} \boldsymbol{K}$, meaning we aim to design a quantum circuit to execute the computation: $\boldsymbol{Z} =\boldsymbol{W}_{V}^{\top} \boldsymbol{X} \boldsymbol{A}$. Here, $\boldsymbol{Z}=\left[\boldsymbol{z}_{1}, \ldots, \boldsymbol{z}_{n}\right] \in \mathbb{R}^{r \times n}$. The matrix elements of $\boldsymbol{A}$ are $\boldsymbol{A}_{uv}=\boldsymbol{q}_u^{\top} \boldsymbol{k}_v=\boldsymbol{x}_u^{\top} \boldsymbol{W}_{Q} \boldsymbol{W}_{K}^{\top} \boldsymbol{x}_v$.}
    \label{attentionfig}
\end{figure}
\subsubsection{Self-Attention(single-head) on Quantum Computer}\label{selfsection}

The "Scaled Dot-Product Attention" defined in Eqn. \ref{attentiondef} can also be written as follows by plugging in Eqn. \ref{quary2} and denoting  $\boldsymbol{A}_0\equiv \operatorname{softmax}\left(\frac{1}{\sqrt{p}} \boldsymbol{Q}^{\top} \boldsymbol{K}\right)$:

\begin{equation}
    	 \boldsymbol{Z}_0
=\boldsymbol{W}_{V}^{\top} \boldsymbol{X} \boldsymbol{A}_0
\label{attentiondef3}	
\end{equation}

Considering it is not straightforward to implement the $\operatorname{softmax}$ function\footnote{Exploring alternatives to the softmax function in attention mechanisms has garnered interest due to the potential for efficiency gains and improved model performance. Research has demonstrated that it's possible to achieve high performance without the need for softmax normalization \cite{lu2021soft,qin2022cosformer}.}using quantum circuit and the scaling will be taken care of in the block-encoding\footnote{Appendix.\ref{blocke} provides a brief introduction of block-encoding.} procedure described later in this section, we aim to implement an alternative version of  $\boldsymbol{A}_0$ denoted as  $\boldsymbol{A}\equiv \boldsymbol{Q}^{\top} \boldsymbol{K}$, that is, we aim to design quantum circuit implementing the following computation:

\begin{equation}
   	 \boldsymbol{Z} =\boldsymbol{W}_{V}^{\top} \boldsymbol{X} \boldsymbol{A},
\label{attentionq}	 
\end{equation}
where $\boldsymbol{Z}=\left[\boldsymbol{z}_{1}, \ldots, \boldsymbol{z}_{n}\right]\in \mathbb{R}^{r \times n} $ .\newline

Note that the matrix elements of $\boldsymbol{A}$ are

\begin{equation}
	\boldsymbol{A}_{uv}=\boldsymbol{q}_u^{\top} \boldsymbol{k}_{v}=\boldsymbol{x}_{u}^{\top} \boldsymbol{W}_{Q} \boldsymbol{W}_{K}^{\top} \boldsymbol{x}_{v}
	\label{matrix-elements }
\end{equation}

The above description of classical attention (the modified version that we aim to implement on the quantum circuit) can be illustrated in Fig.\ref{attentionfig}. Next, we present its quantum implementation.\newline

\begin{figure}[h!]
\centering
 \includegraphics[width=\linewidth]{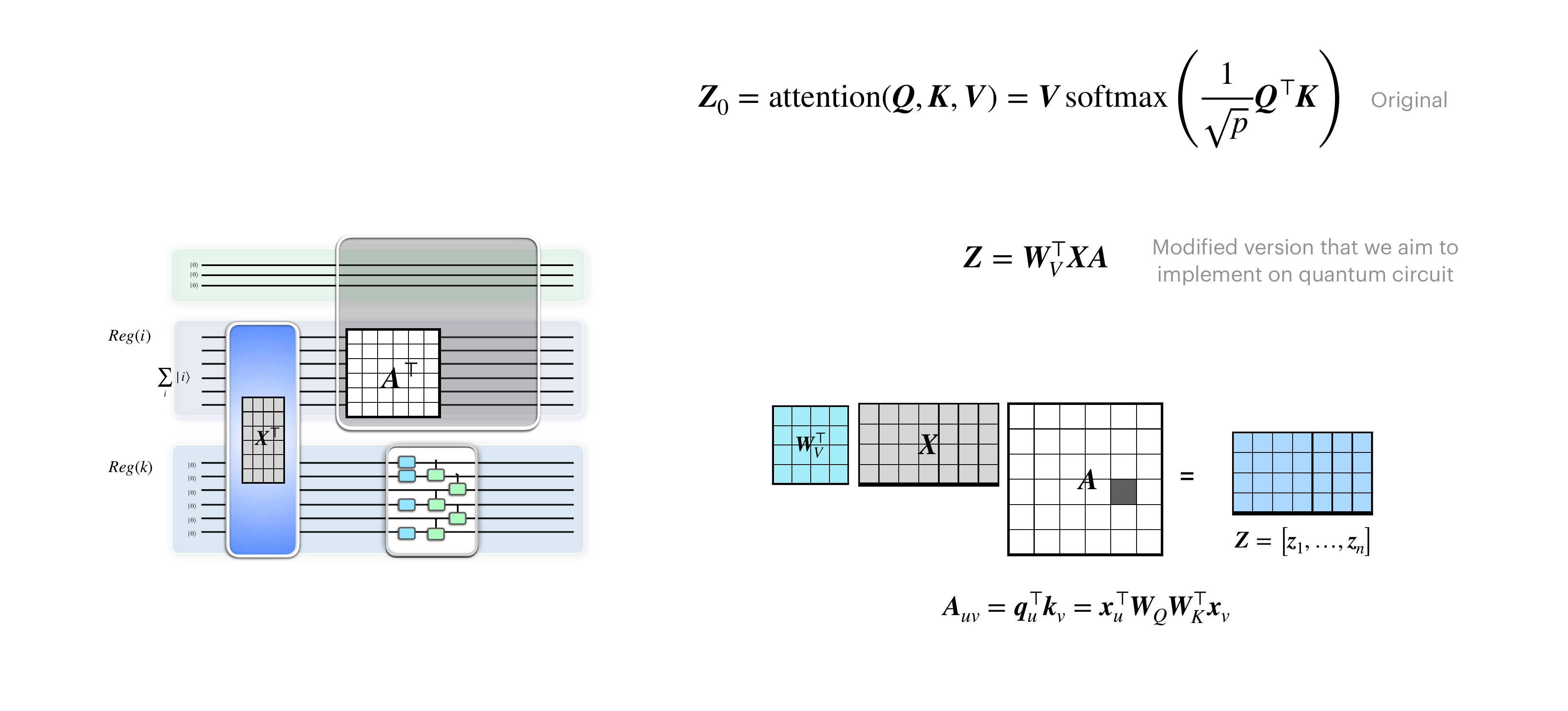}
   \caption{\textit{Self-Attention on Quantum Computer} The right side of the figure describes classical attention (the modified version that we aim to implement on a quantum circuit), and the left side of the figure depicts its quantum implementation. On the quantum circuit, the input encoding is represented by the blue box, as described in Section \ref{input}. The attention function can be implemented by applying the block-encoding of $\boldsymbol{A}^{\top}$ and a parameterized quantum circuit for $\boldsymbol{W}_{V}^{\top} $ on the two quantum registers $Reg(i)$ and $Reg(k)$ respectively.}
    \label{qattention1}
\end{figure}

On a quantum circuit, after the input encoding described in \ref{input}, the attention function can be implemented by applying the block-encoding of $\boldsymbol{A}^{\top}$ and a parameterized quantum circuit\footnote{Appendix.\ref{para} provides a brief introduction of using parametrized quantum circuit for implementing trainable linear transformations.} for $\boldsymbol{W}_{V}^{\top}$ on the two quantum registers $Reg(i)$ and $Reg(k)$ respectively, as depicted in Fig.~\ref{qattention1}. \newline

This can be proven as follows:\newline

Starting from Eq.~\ref{attentionq}, we have
\begin{equation*}
    \boldsymbol{Z} = \boldsymbol{W}_{V}^{\top} \boldsymbol{X} \boldsymbol{A},
\end{equation*}

Utilizing the formula $\textit{vec}(ABC) = (C^{\top} \otimes A)\textit{vec}(B)$, we obtain

\begin{equation}
    \textit{vec}(\boldsymbol{Z}) = \textit{vec}(\boldsymbol{W}_{V}^{\top} \boldsymbol{X} \boldsymbol{A})
    = (\boldsymbol{A}^{\top} \otimes \boldsymbol{W}_{V}^{\top})\textit{vec}(\boldsymbol{X}).
    \label{vec}
\end{equation}

For a matrix $M$, defining vectors $\vec{\psi_{M}} = \textit{vec}(M)$ transforms Eq.~\ref{vec} into

\begin{equation}
    \vec{\psi_{\boldsymbol{Z}}} = (\boldsymbol{A}^{\top} \otimes \boldsymbol{W}_{V}^{\top}) \vec{\psi_{\boldsymbol{X}}}.
    \label{vec2}
\end{equation}

Recall Eq.~\ref{psix}:

\begin{equation*}
    \left| \psi_{\boldsymbol{X}} \right\rangle = \sum_{i=1}^{n} |i\rangle\ket{\boldsymbol{x}_{i}},
\end{equation*}

where writing the quantum states in Eq.~\ref{psix} as vectors yields
\begin{equation}
    \vec{\psi_{\boldsymbol{X}}} = \left| \psi_{\boldsymbol{X}} \right\rangle.
    \label{tensor}
\end{equation}

Correspondingly, we have
\begin{equation}
    \vec{\psi_{\boldsymbol{Z}}} = \left| \psi_{\boldsymbol{Z}} \right\rangle,
    \label{tensorz}
\end{equation}

by defining $ \left| \psi_{\boldsymbol{Z}} \right\rangle := \sum_{i=1}^{n} |i\rangle\ket{\boldsymbol{z}_{i}},$ where $\ket{\boldsymbol{z}_{i}} := \sum_{k=1}^{d} \boldsymbol{z}^{(k)}_{i} |k\rangle$ is the amplitude encoding of the vector $\boldsymbol{z}_{i}$ with its $k$-th elements denoted by $\boldsymbol{z}^{(k)}_{i}$.\newline

From Eq.~\ref{tensorz} to Eq.~\ref{vec2}, we derive
\begin{equation}
    \left| \psi_{\boldsymbol{Z}} \right\rangle = (\boldsymbol{A}^{\top} \otimes \boldsymbol{W}_{V}^{\top}) \left| \psi_{\boldsymbol{X}} \right\rangle.
    \label{v}
\end{equation}

This corresponds to the action of the quantum circuit for self-attention, which can be represented as

\begin{equation}
    \left| \psi_{\boldsymbol{Z}} \right\rangle \otimes \ket{0} + \ldots = (U_{\boldsymbol{A}^{\top}} \otimes U_{\boldsymbol{W}_{V}^{\top}}) (\left| \psi_{\boldsymbol{X}} \right\rangle \otimes \ket{0}),
    \label{v2}
\end{equation}

where $U_{\boldsymbol{A}^{\top}}$ corresponds to applying the block-encoding of $\boldsymbol{A}^{\top}$, and $U_{\boldsymbol{W}_{V}^{\top}}$ is a parameterized quantum circuit implementing $\boldsymbol{W}_{V}^{\top}$ on the two quantum registers $Reg(i)$ and $Reg(k)$ respectively. The term "$+\ldots$"\footnote{Throughout this paper, the terms "$+\ldots$" in the quantum states are consistently used as defined here: "$\ldots$" represents a quantum state that is orthogonal to the state before the "$+$" sign.} indicates that upon post-selecting, we can obtain the desired state $\left| \psi_{\boldsymbol{Z}} \right\rangle = \textit{vec}(\boldsymbol{Z})$.\newline

The block-encoding of $\boldsymbol{A}^{\top}$ can be constructed using the following lemma from Ref.~\cite{nguyen2022block}.\newline

\textbf{Lemma 3.2 from Ref.~\cite{nguyen2022block} } (\textit{Naive block-encoding of dense matrices with oracle access}). Let \( A \in \mathbb{C}^{N \times N} \) (where \( N = 2^s \)) with $a_{ij}$ being its elements and let \( \hat{a} \geq \max_{i,j} |a_{ij}| \). Suppose the following oracle is provided
\[
O_A : |i\rangle |j\rangle |0\rangle^{\otimes b} \rightarrow |i\rangle |j\rangle |\tilde{a}_{ij}\rangle,
\]
where \( 0 \leq i,j < N \) and \( \tilde{a}_{ij} \) is the (exact) b-qubit description of \( a_{ij} / \hat{a} \). Then one can implement a \( (N\hat{a}, s + 1, \epsilon) \)-block-encoding of \( A \) with two uses of \( O_A \), \( O(\text{polylog}(\hat{a}N / \epsilon)) \) one- and two-qubit gates and \( O(b, \text{polylog}(\hat{a}N / \epsilon)) \) extra qubits (which are discarded before the post-selection step).\newline

Below we explain how the block-encoding of $\boldsymbol{A}^{\top}\in \mathbb{R}^{n \times n}$ can be constructed using the above lemma: \newline

From Eqn.\ref{matrix-elements }, the matrix elements of $\boldsymbol{A}^{\top}$ (which we denote as $\Lambda_{ij})$ are

\begin{equation}
	 \Lambda_{ij}:={\boldsymbol{A}^{\top}}_{ij}=\boldsymbol{x}_{j}^{\top} \boldsymbol{W}_{Q} \boldsymbol{W}_{K}^{\top} \boldsymbol{x}_{i}
	 \label{elements2}
\end{equation}

let \( \hat{\Lambda} \geq \max_{i,j} |\Lambda_{ij}| \) and \( \tilde{\Lambda}_{ij} \) is defined to be the (exact) b-qubit description of \( \Lambda_{ij} / \hat{\Lambda} \).\newline

According to the above lemma, we need the following oracle to the block-encoding of $\boldsymbol{A}^{\top}$

\begin{equation}
	O_{\boldsymbol{A}^{\top}} : |i\rangle |j\rangle |0\rangle^{\otimes b} \rightarrow |i\rangle |j\rangle |\tilde{\Lambda}_{ij}\rangle,
	\end{equation}
where \( 0 \leq i,j < n \).\newline

This oracle $O_{\boldsymbol{A}^{\top}}$ can be constructed in the same way as constructing $ O_{\textsf{attention}}$ (described in Appendix \ref{attentionsec}) defined in Eqn.\ref{attentionoracle} with the attention score as Eqn.\ref{attention-score }. Therefore, substituting $A$ in the above lemma with $\boldsymbol{A}^{\top}$, we can implement the block-encoding of $\boldsymbol{A}^{\top}$ with two uses of $ O_{\boldsymbol{A}^{\top}}$, \( O(\text{polylog}(\hat{\Lambda}n / \epsilon)) \) one- and two-qubit gates.

\subsubsection{Multihead-Attention on Quantum computer}\label{sectionmulti}
In the multihead attention module, We have $H$  set of queries, values, and keys as \cite{ghojogh2020attention}:

$$
\begin{aligned}
& \mathbb{R}^{p \times n} \ni \boldsymbol{Q}_{h}=\boldsymbol{W}_{Q,h}^{\top} \boldsymbol{X}, \quad \forall h \in\{1, \ldots, H\}, \\
& \mathbb{R}^{p \times n} \ni \boldsymbol{V}_{h}=\boldsymbol{W}_{V,h}^{\top} \boldsymbol{X}, \quad \forall h \in\{1, \ldots, H\}, \\
& \mathbb{R}^{r \times n} \ni \boldsymbol{K}_{h}=\boldsymbol{W}_{K,h}^{\top} \boldsymbol{X}, \quad \forall h \in\{1, \ldots, H\} .
\end{aligned}
$$

Then, the scaled dot product attention is applied to generate the $H$ output  $\left\{\boldsymbol{Z}_{h}\right\}_{h=1}^{H}$, in accordance with Eqn.\ref{attentionq}:

\begin{equation}
   	 \boldsymbol{Z}_h =\boldsymbol{W}_{V,h}^{\top} \boldsymbol{X} \boldsymbol{A}_h,
\label{attentiondef2}	 
\end{equation}

where  $\boldsymbol{A}_h\equiv \boldsymbol{Q}_h^{\top} \boldsymbol{K}_h$, and $\boldsymbol{Z}_h=\left[\boldsymbol{z}_{1,h}, \ldots, \boldsymbol{z}_{n,h}\right]\in \mathbb{R}^{r \times n} $. \newline

The outputs are concatenated over different heads as

\begin{equation}
	\boldsymbol{Z}_{\textsf{Multi-heads}}=\left[\parallel_{h=1}^{H}\boldsymbol{z}_{1,h}, \parallel_{h=1}^{H}\boldsymbol{z}_{2,h},\ldots, \parallel_{h=1}^{H}\boldsymbol{z}_{n,h},\right]\in \mathbb{R}^{rH \times n}
\end{equation}

where $\parallel$ represents concatenation \cite{velickovic2017graph}.
 Then, by a linear projection $\boldsymbol{W}_{O}^{\top}$, the total attention value is obtained:

\begin{equation}
\boldsymbol{z}_i^{\textsf{Total}}:=\boldsymbol{W}_{O}^{\top} \parallel_{h=1}^{H}\boldsymbol{z}_{i,h}
\end{equation}

$$
\boldsymbol{Z}^{\textsf{Total}}:=\boldsymbol{W}_{O}^{\top} \boldsymbol{Z}_{\textsf{Multi-heads}}$$

and $\boldsymbol{Z}^{\textsf{Total}}=\left[\boldsymbol{z}^{\textsf{Total}}_{1}, \ldots, \boldsymbol{z}^{\textsf{Total}}_{n}\right]\in \mathbb{R}^{rH \times n} $.\newline

\begin{figure}[h!]
    \centering
    \includegraphics[width=\linewidth]{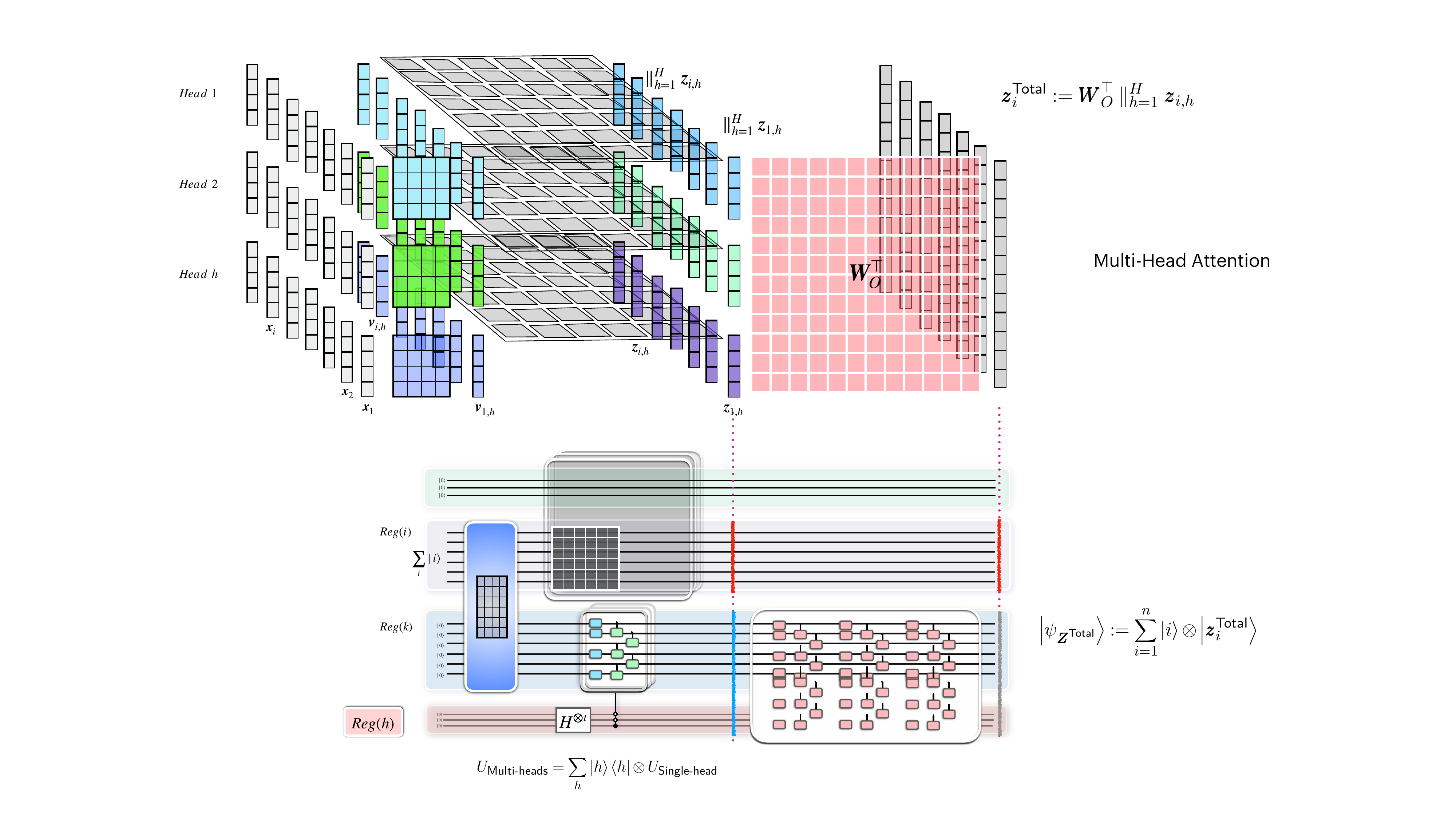}
   \caption{\textit{Multihead-Attention on Quantum Computer}. The upper part of the figure provides the illustration of classical Multihead-Attention, and the lower part of the figure depicts its quantum implementation. 
   In the multihead attention module, We have $H$  set of queries, values, and keys as:$\mathbb{R}^{p \times n} \ni \boldsymbol{Q}_{h}=\boldsymbol{W}_{Q,h}^{\top} \boldsymbol{X}, \quad \forall h \in\{1, \ldots, H\}, \mathbb{R}^{p \times n} \ni \boldsymbol{V}_{h}=\boldsymbol{W}_{V,h}^{\top} \boldsymbol{X}, \quad \forall h \in\{1, \ldots, H\},  \mathbb{R}^{r \times n} \ni \boldsymbol{K}_{h}=\boldsymbol{W}_{K,h}^{\top} \boldsymbol{X}, \quad \forall h \in\{1, \ldots, H\} $ Then, the scaled dot product attention are applied to generate the $H$ output  $\left\{\boldsymbol{Z}_{h}\right\}_{h=1}^{H}$ and $\boldsymbol{Z}_h=\left[\boldsymbol{z}_{1,h}, \ldots, \boldsymbol{z}_{n,h}\right]\in \mathbb{R}^{r \times n}$. The outputs are concatenated over different heads as $
	\boldsymbol{Z}_{\textsf{Multi-heads}}=\left[\parallel_{h=1}^{H}\boldsymbol{z}_{1,h}, \parallel_{h=1}^{H}\boldsymbol{z}_{2,h},\ldots, \parallel_{h=1}^{H}\boldsymbol{z}_{n,h},\right]\in \mathbb{R}^{rH \times n}$, where $\parallel$ represents concatenation \cite{velickovic2017graph}.
 Then, by a linear projection $\boldsymbol{W}_{O}^{\top}$, the total attention value is obtained: $	\boldsymbol{z}_i^{\textsf{Total}}:=\boldsymbol{W}_{O}^{\top} \parallel_{h=1}^{H}\boldsymbol{z}_{i,h}$,$
\boldsymbol{Z}^{\textsf{Total}}:=\boldsymbol{W}_{O}^{\top} \boldsymbol{Z}_{\textsf{Multi-heads}}$ and $\boldsymbol{Z}^{\textsf{Total}}=\left[\boldsymbol{z}^{\textsf{Total}}_{1}, \ldots, \boldsymbol{z}^{\textsf{Total}}_{n}\right]\in \mathbb{R}^{rH \times n} $. On the quantum circuit, the input encoding is represented by the blue box, as described in Section \ref{input}. The multi-head attention can be implemented by applying a multi-controlled unitary defined as $ U_{\textsf{Multi-heads}} =\sum_h \ket{h}\bra{h}\otimes U_{\textsf{Single-head}} $ where $U_{\textsf{Single-head}}= (U_{\boldsymbol{A}_h^{\top}}\otimes U_{\boldsymbol{W}_{V,h}^{\top}})$ representing the block-encoding of $\boldsymbol{A}_h^{\top}$ and a parameterized quantum circuit for $\boldsymbol{W}_{V,h}^{\top}$ for each head.}
    \label{Multihead-Attention}
\end{figure}

The above description of classical multihead attention can be illustrated in Fig.\ref{Multihead-Attention}. Next, we present its quantum implementation. \newline

On a quantum circuit, we aim to obtain the following quantum state,

\begin{equation}
	\left|\psi_{\boldsymbol{Z}^{\textsf{Total}}}\right\rangle:=\sum_{i=1}^{n} |i\rangle\otimes \ket{\boldsymbol{z}^{\textsf{Total}}_{i}}
	\label{desire1}
	\end{equation}

where $\ket{\boldsymbol{z}^{\textsf{Total}}_{i}}$ is the amplitude encoding of the vector $\boldsymbol{z}^{\textsf{Total}}_{i}$. This can be achieved via the quantum circuit depicted in Fig.\ref{Multihead-Attention} in which the additional register $Reg(h)$ is hosting index $h$, and the multi-controlled unitary is defined as

\begin{equation}
  U_{\textsf{Multi-heads}} =\sum_h \ket{h}\bra{h}\otimes U_{\textsf{Single-head}} \end{equation}

with $U_{\textsf{Single-head}}= (U_{\boldsymbol{A}_h^{\top}}\otimes U_{\boldsymbol{W}_{V,h}^{\top}})$ representing the block-encoding of $\boldsymbol{A}_h^{\top}$ and a parameterized quantum circuit for $\boldsymbol{W}_{V,h}^{\top}$ for each head.\newline

For a single head, from Eq.~\ref{v2}, when $U_{\textsf{Single-head}}$ acts on the state $\left|\psi_{\boldsymbol{X}}\right\rangle \otimes \ket{0}$, the outcome state is

\begin{equation}
  U_{\textsf{Single-head}}\left(\left|\psi_{\boldsymbol{X}}\right\rangle \otimes \ket{0}\right) = \left|\psi_{\boldsymbol{Z}_h}\right\rangle\otimes \ket{0} + \ldots
\end{equation}
where $\left|\psi_{\boldsymbol{Z}_h}\right\rangle := \sum_{i=1}^{n} |i\rangle \otimes \ket{\boldsymbol{z}_{i,h}}$ and $\ket{\boldsymbol{z}_{i,h}}$ is the amplitude encoding of the vector $\boldsymbol{z}_{i,h}$.\newline

When $U_{\textsf{Multi-heads}}$ acts on the state prepared as $\sum_h \ket{h} \otimes \left(\left|\psi_{\boldsymbol{X}}\right\rangle \otimes \ket{0}\right)$ by the blue box and Hadamard gates on $Reg(h)$, the outcome state is

\begin{equation}
  \left|\psi_{\textsf{Multi-heads}}\right\rangle = U_{\textsf{Multi-heads}}\left(\sum_h \ket{h} \otimes \left(\left|\psi_{\boldsymbol{X}}\right\rangle \otimes \ket{0}\right)\right) = \sum_h \ket{h} \otimes \left(\left|\psi_{\boldsymbol{Z}_h}\right\rangle \otimes \ket{0} + \ldots\right)
\end{equation}

upon post-selecting, we obtain the state

\begin{align}
  \left|\psi_{\textsf{Multi-heads}}\right\rangle = \sum_{i=1}^{n} |i\rangle\otimes \sum_h \ket{h}\ket{\boldsymbol{z}_{i,h}}	
\\=\sum_{i=1}^{n} |i\rangle\otimes\ket{\parallel_{h=1}^{H}\boldsymbol{z}_{i,h}}
\end{align}

where $\ket{\parallel_{h=1}^{H}\boldsymbol{z}_{i,h}}$ is the amplitude encoding of the vector $\parallel_{h=1}^{H}\boldsymbol{z}_{i,h}$.\newline

Applying  a parameterized quantum circuit implementing $\boldsymbol{W}_{O}^{\top}$ on $Reg(k)$ and $Reg(h)$, which act as $U_{\boldsymbol{W}_{O}^{\top}} \ket{\parallel_{h=1}^{H}\boldsymbol{z}_{i,h}}=\ket{\boldsymbol{z}^{\textsf{Total}}_{i}}$, obtain

\begin{align}
 U_{\boldsymbol{W}_{O}^{\top}} \left|\psi_{\textsf{Multi-heads}}\right\rangle = \sum_{i=1}^{n} |i\rangle\otimes U_{\boldsymbol{W}_{O}^{\top}} \ket{\parallel_{h=1}^{H}\boldsymbol{z}_{i,h}}\\=\sum_{i=1}^{n} |i\rangle\otimes \ket{\boldsymbol{z}^{\textsf{Total}}_{i}}\\=\left|\psi_{\boldsymbol{Z^{\textsf{Total}}}}\right\rangle
\end{align}

which is the desired state in Eqn.\ref{desire1}.

\subsection{Residual-connection on Quantum computer}

After the multihead attention module, the input to the transformer block $\boldsymbol{x}_{i}$ and the total attention value $\boldsymbol{z}^{\textsf{Total}}_{i}$ are added (often referred as "residual-connection" introduced by ResNet \cite{he2016deep}):

\begin{equation}
	\boldsymbol{z}'_{i}:=\boldsymbol{z}^{\textsf{Total}}_{i}+\operatorname{concat}(\boldsymbol{x}_{i},\boldsymbol{x}_{i},...\boldsymbol{x}_{i})\in \mathbb{R}^{rH} 
\end{equation}

where $\operatorname{concat}(\boldsymbol{x}_{i},\boldsymbol{x}_{i},...\boldsymbol{x}_{i})$ represents the concatenation of $H$ identical vectors\footnote{Note that this is a bit different from the standard classical residual connection.} $\boldsymbol{x}_{i}$, For later use, define $\ket{\operatorname{concat}(\boldsymbol{x}_{i},\boldsymbol{x}_{i},...\boldsymbol{x}_{i})}$ as the amplitude encoding of the vector $\operatorname{concat}(\boldsymbol{x}_{i},\boldsymbol{x}_{i},...\boldsymbol{x}_{i})$ and $\boldsymbol{Z'}:=\left[\boldsymbol{z'}_{1}, \ldots, \boldsymbol{z'}_{n}\right]\in \mathbb{R}^{rH \times n} $.\newline

\begin{figure}[h!]
    \centering
    \includegraphics[width=\linewidth]{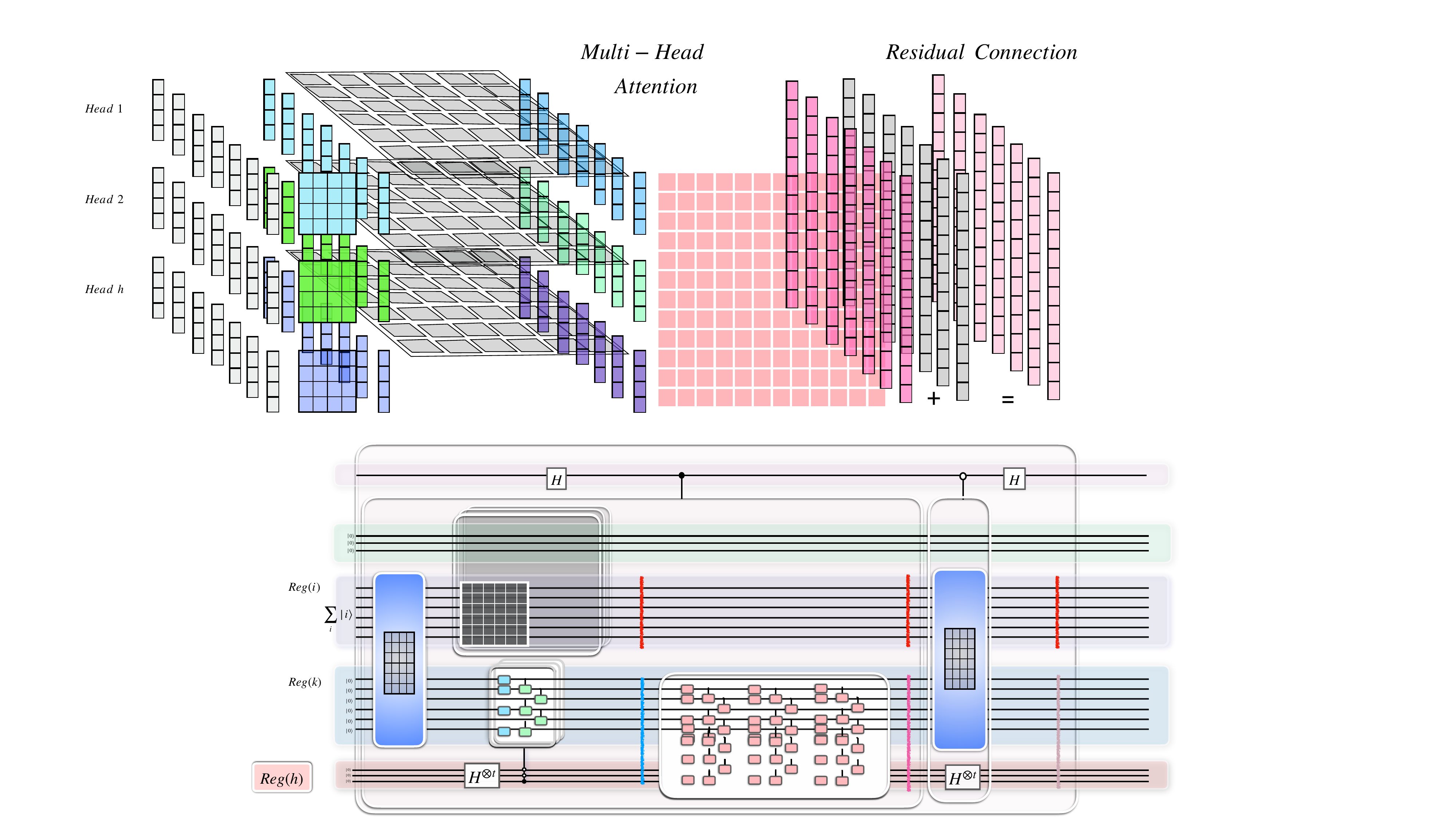}
   \caption{\textit{Residual-connection on Quantum computer} The upper part of the figure provides the illustration of classical Multihead-Attention followed by Residual-connection, the lower part of the figure depicts their quantum implementation. After the multihead attention module, the data (containing positional encoding) $\boldsymbol{x}_{i}$ and the total attention value $\boldsymbol{z}^{\textsf{Total}}_{i}$ are added (often referred as "residual-connection" introduced by ResNet \cite{he2016deep}):$\boldsymbol{z}'_{i}:=\boldsymbol{z}^{\textsf{Total}}_{i}+\operatorname{concat}(\boldsymbol{x}_{i},\boldsymbol{x}_{i},...\boldsymbol{x}_{i})\in \mathbb{R}^{rH} $. The quantum circuit in this figure generates the following quantum state $\left|\psi_{\boldsymbol{Z'}}\right\rangle:=\sum_{i=1}^{n} |i\rangle\otimes \ket{\boldsymbol{z'}_{i}}
	$ where $\ket{\boldsymbol{z'}_{i}}$ is the amplitude encoding of the vector $\boldsymbol{z'}_{i}$.  The circuit implements Linear Combination of Unitaries of two operators grouped in the two transparent boxes controlled by the top ancillary qubit. }
    \label{Residue}
\end{figure}

On the quantum circuit, we aim to obtain the following quantum state

\begin{equation}
	\left|\psi_{\boldsymbol{Z'}}\right\rangle:=\sum_{i=1}^{n} |i\rangle\otimes \ket{\boldsymbol{z'}_{i}}
	\label{desire2}	\end{equation}

where $\ket{\boldsymbol{z'}_{i}}$ is the amplitude encoding of the vector $\boldsymbol{z'}_{i}$.  This can be achieved via the quantum circuit depicted in Fig.\ref{Residue}.

The first transparent box controlled by the top ancillary qubit in Fig.\ref{Residue} implements $U_{\boldsymbol{Z^{\textsf{Total}}}}$ which acts as

\begin{equation}
U_{\boldsymbol{Z^{\textsf{Total}}}}:|i\rangle\otimes \ket{0} \to|i\rangle\otimes \ket{\boldsymbol{z}^{\textsf{Total}}_{i}}
  \end{equation}

The blue box represents  the data encoding (containing positional encoding) which acts as
\begin{equation}
U_{\boldsymbol{X}}: |i\rangle\otimes \ket{0}
 \to |i\rangle\otimes \ket{\boldsymbol{x}_i}
  \end{equation}

Including the Hadamard gates, the second transparent box controlled by the top ancillary qubit acts on the input state as
\begin{equation}
U_{\boldsymbol{X}}\otimes H^{\otimes t} (\sum_{i=1}^{n} |i\rangle\otimes\ket{0} \otimes\ket{0} )=\sum_{i=1}^{n} |i\rangle\otimes \ket{\boldsymbol{x}_{i}}\otimes\sum_h \ket{h} =\left|\psi_{\boldsymbol{X}}\right\rangle\otimes\sum_h \ket{h} 
\end{equation}

where $t=\log_2(H)$, and the $H$ here is the number of "heads" in the multi-head attention defined in Section \ref{sectionmulti}. \newline

The circuit in Fig.\ref{Residue} implements a linear combination of two unitaries as in the two transparent boxes controlled by the top ancillary qubit; it generates the state $(\left|\psi_{\boldsymbol{Z^{\textsf{Total}}}}\right\rangle +\left|\psi_{\boldsymbol{X}}\right\rangle\otimes\sum_h \ket{h}) \ket{0}+\ldots$ in which $(\left|\psi_{\boldsymbol{Z^{\textsf{Total}}}}\right\rangle +\left|\psi_{\boldsymbol{X}}\right\rangle\otimes\sum_h \ket{h})$ can be rewritten as follows:

\begin{equation}
\begin{aligned}
\left|\psi_{\boldsymbol{Z}^{\textsf{Total}}}\right\rangle +\left|\psi_{\boldsymbol{X}}\right\rangle \otimes\sum_h \left| h \right\rangle &= \sum_{i=1}^{n} \left| i \right\rangle \otimes \left| \boldsymbol{z}^{\textsf{Total}}_{i} \right\rangle + \sum_{i=1}^{n} \left| i \right\rangle \otimes \left| \boldsymbol{x}_{i} \right\rangle \otimes\sum_h \left| h \right\rangle \\
&= \sum_{i=1}^{n} \left| i \right\rangle \otimes \left( \left| \boldsymbol{z}^{\textsf{Total}}_{i} \right\rangle + \left| \boldsymbol{x}_{i} \right\rangle \otimes\sum_h \left| h \right\rangle \right) \\
&= \sum_{i=1}^{n} \left| i \right\rangle \otimes \left( \left| \boldsymbol{z}^{\textsf{Total}}_{i} \right\rangle + \sum_h \left| h \right\rangle \left| \boldsymbol{x}_{i} \right\rangle \right) \\
&= \sum_{i=1}^{n} \left| i \right\rangle \otimes \left( \left| \boldsymbol{z}^{\textsf{Total}}_{i} \right\rangle + \left| \operatorname{concat}(\boldsymbol{x}_{i}, \boldsymbol{x}_{i}, \ldots, \boldsymbol{x}_{i}) \right\rangle \right) \\
&= \sum_{i=1}^{n} \left| i \right\rangle \otimes \left| \boldsymbol{z}'_{i} \right\rangle = \left| \psi_{\boldsymbol{Z'}} \right\rangle.
\end{aligned}
\end{equation}

which is the desired state in Eqn.\ref{desire2}.\newline

In our quantum implementation of GPT, the layer normalization procedure is omitted, our quantum approach leverages the inherent unitarity of quantum state evolutions to achieve normalization. 

\subsection{Feed-Forward Network on Quantum computer}\label{ffn}
After the multihead attention module and residual connection, a position-wise Feed-Forward Network(FFN) is applied\cite{ghojogh2020attention}. The  $\mathrm{FFN}$ is a fully connected feed-forward module that operates separately and identically on each $\boldsymbol{z}'_{i}$:

$$
\operatorname{FFN}\left(\boldsymbol{z}'_{i}\right)={{\mathbf{W}^2}}^{\top}\operatorname{ReLU}\left({\mathbf{W}^1}^{\top}\boldsymbol{z}'_{i}+\mathbf{b}^1\right) +\mathbf{b}^2,
$$
where $\mathbf{W}^1 \in \mathbb{R}^{rH \times d_{ff}}, \mathbf{W}^2 \in$ $\mathbb{R}^{d_{ff} \times rH}, \mathbf{b}^1 \in \mathbb{R}^{d_{ff}}, \mathbf{b}^2 \in \mathbb{R}^{rH}$ are trainable parameters, $d_{ff}$ is the intermediate dimension of the FFN. For simplicity, we omit $\mathbf{b}^1, \mathbf{b}^2 $ in the following discussion. Similar to how we defined $\boldsymbol{X}:=\left[\boldsymbol{x}_{1}, \ldots, \boldsymbol{x}_{n}\right] $ before, we can write $\mathbf{W}^1 =\left[\boldsymbol{w}_{1}, \ldots, \boldsymbol{w}_{m},\ldots,\boldsymbol{w}_{d_{ff}}\right] $ where $\boldsymbol{w}_{m} \in \mathbb{R}^{rH}$. \newline

Denote $\boldsymbol{y}_{i}:={\mathbf{W}^1}^{\top}\boldsymbol{z}'_{i}\in \mathbb{R}^{ d_{ff}}$, we have its elements as

\begin{equation}
	\boldsymbol{y}_{i}^{(m)}=\boldsymbol{z}'_{i} \cdot	\boldsymbol{w}_{m},        \forall m \in\{1, \ldots, d_{ff}\}
	\end{equation}

\begin{figure}[h!]
    \centering
    \includegraphics[width=\linewidth]{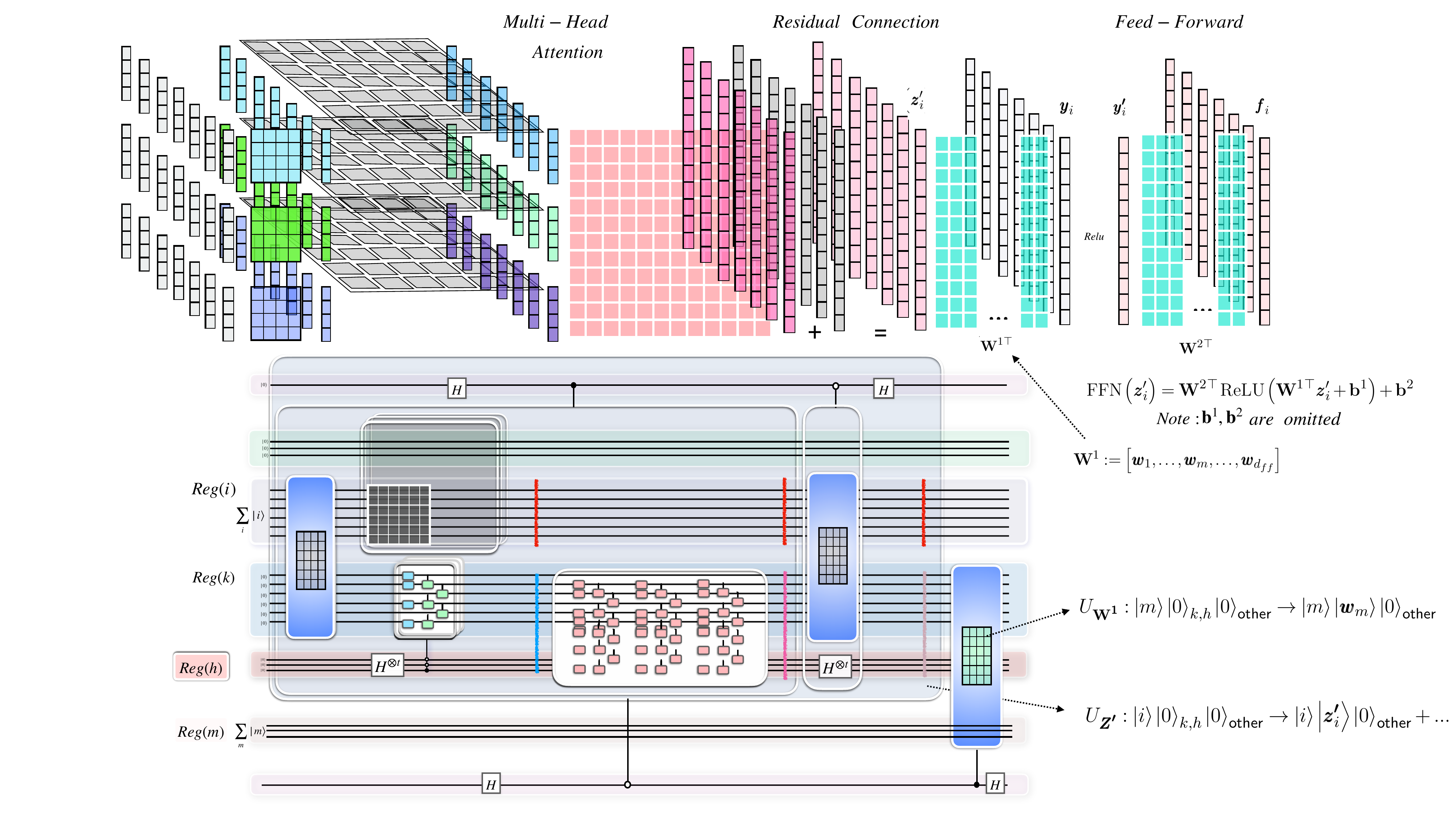}
   \caption{\textit{Feed-Forward Network on Quantum computer 1} The upper part of the figure provides the illustration of classical Multihead-Attention followed by residual connection and Feed-Forward Network, and the lower part of the figure depicts their quantum implementation. After the multihead attention module and residual connection, a position-wise Feed-Forward Network(FFN) is applied. The  $\mathrm{FFN}$ is a fully connected feed-forward module that operates separately and identically on each $\boldsymbol{z}'_{i}$: ${\mathbf{W}^2}^{\top}\operatorname{ReLU}\left({\mathbf{W}^1}^{\top}\boldsymbol{z}'_{i}+\mathbf{b}^1\right) +\mathbf{b}^2$,  we can write $\mathbf{W}^1 :=\left[\boldsymbol{w}_{1}, \ldots, \boldsymbol{w}_{m},\ldots,\boldsymbol{w}_{d_{ff}}\right] $ where $\boldsymbol{w}_{m} \in \mathbb{R}^{rH}$,$d_{ff}$ is the intermediate dimension $d_{ff}$ of the FFN. Denote $\boldsymbol{y}_{i}:={\mathbf{W}^1}^{\top}\boldsymbol{z}'_{i}$, we have its elements as $\boldsymbol{y}_{i}^{(m)}=\boldsymbol{z}'_{i} \cdot	\boldsymbol{w}_{m}$. Recall we created state on registered $Reg(i), Reg(k), Reg(h)$ and ancillas: $\left|\psi_{\boldsymbol{Z'}}\right\rangle =\sum_{i=1}^{n} |i\rangle\otimes \ket{\boldsymbol{z}'_{i}}\otimes \ket{0}+...$, by the unitary circled in the overall transparent box in this figure, denoted as $U_{\boldsymbol{Z'}}$, which act as $U_{\boldsymbol{Z'}}: |i\rangle \ket{0}_{k,h} \ket{0}_{\textsf{other}}\to |i\rangle\ket{\boldsymbol{z'}_{i}}\ket{0}_{\textsf{other}}+..., \forall i \in\{1,\cdots n\}.$ For implementing $\mathbf{W}^1$, we can create a trainable unitary $U_{\boldsymbol{\mathbf{W}^1}}: |m\rangle \ket{0}_{k,h}\ket{0}_{\textsf{other}}\to |m\rangle\ket{\boldsymbol{w}_{m}}\ket{0}_{\textsf{other}}, \forall m \in\{1, \ldots, d_{ff}\}.$ with $|\boldsymbol{w}_{m}\rangle$ on $Reg(k), Reg(h)$ and $ |m\rangle$ on an additional registered $Reg(m)$. $U_{\boldsymbol{\mathbf{W}^1}}$, depicted as the blue box with a green centre in this figure, can be implemented as a series of controlled parameterised quantum circuits as $U_{\boldsymbol{\mathbf{W}^1}}=\sum_m \ket{m}\bra{m}U_m$ where each $U_m$, acting as $U_m:\ket{0}_{k,h}\to \ket{\boldsymbol{w}_{m}}$, is a parameterised quantum circuit.$\boldsymbol{y}_{i}^{(m)}=\langle\boldsymbol{z}'_{i} |\boldsymbol{w}_{m}\rangle$ can be evaluated using Parallel Swap test for each $\boldsymbol{z}'_{i}$ and $\boldsymbol{w}_{m}$, via the quantum circuit depicted in this figure .}
    \label{feed1}
\end{figure}

Next, we present a quantum implementation of the FFN module similar to the approach proposed in Ref.~\cite{allcock2020quantum}.\newline

Recall the unitary circled in the overall transparent box in Fig.\ref{Residue}, denoted as $U_{\boldsymbol{Z'}}$, act as
\begin{equation}
U_{\boldsymbol{Z'}}: |i\rangle \ket{0}_{k,h} \ket{0}_{\textsf{other}}\to |i\rangle\ket{\boldsymbol{z'}_{i}}\ket{0}_{\textsf{other}}+..., \forall i \in\{1,\cdots n\}.
\label{Ux}
\end{equation}
where $\ket{0}_{k,h}$ represents the "all-zero" state \footnote{in this paper, the 'all-zero' state refers to all the relevant qubits being in the $\ket{0}$ state.} of the qubits in $Reg(k)$ and $Reg(h)$, $\ket{0}_{\textsf{other}}$ represents the "all-zero" state of the ancillary qubits(qubits that are not included in $Reg(i)$,$Reg(k)$,$Reg(h)$ in Fig.\ref{Residue}).\newline

For implementing $\mathbf{W}^1 :=\left[\boldsymbol{w}_{1}, \ldots, \boldsymbol{w}_{m},\ldots,\boldsymbol{w}_{d_{ff}}\right] $, we can create a trainable unitary
\begin{equation}
U_{\boldsymbol{\mathbf{W}^1}}: |m\rangle \ket{0}_{k,h}\to |m\rangle\ket{\boldsymbol{w}_{m}}, \forall m \in\{1, \ldots, d_{ff}\}.
\label{Ux2}
\end{equation}

with $ |\boldsymbol{w}_{m}\rangle $ on $Reg(k), Reg(h)$ and $ |m\rangle$ on an additional registered $Reg(m)$. \newline

%This trainable unitary $U_{\boldsymbol{\mathbf{W}^1}}$ can be implemented as a series of controlled parameterised quantum circuits as $U_{\boldsymbol{\mathbf{W}^1}}=\sum_m \ket{m}\bra{m}U_m$ where each $U_m$, acting as $U_m:\ket{0}_{k,h}\to \ket{\boldsymbol{w}_{m}}$, is a parameterised quantum circuit.\newline

Notice that 
\begin{equation}
	\boldsymbol{y}_{i}^{(m)}=\langle\boldsymbol{z}'_{i} |\boldsymbol{w}_{m}\rangle ,        \forall m \in\{1, \ldots, d_{ff}\}
	\end{equation}

This can be evaluated using parallel swap test for each $\boldsymbol{z}'_{i}\in \mathbb{R}^{rH}$ and $\boldsymbol{w}_{m}\in \mathbb{R}^{rH}$, via the quantum circuit depicted in Fig.\ref{feed1}.\newline

The input state to the circuit is

\begin{equation}
\left|\Psi_{0}\right\rangle=\sum_{i}\sum_{m}|i\rangle|m\rangle \ket{0}_{k,h}   \ket{0}_{\textsf{other}}\ket{0}
\end{equation}

For each branch $|i\rangle|m\rangle \ket{0}_{k,h}   \ket{0}_{\textsf{other}}\ket{0}$, applying a Hadamard gate on the bottom ancillary qubit, and controlled $U_{\boldsymbol{Z'}}$, $U_{\boldsymbol{\mathbf{W}^1}}$ we obtain

\begin{equation}
|i\rangle|m\rangle (\ket{\boldsymbol{z'}_{i}}\ket{0}_{\textsf{other}}+...)\ket{0}+|i\rangle|m\rangle \ket{\boldsymbol{w}_{m}}   \ket{0}_{\textsf{other}}\ket{1}
\end{equation}

Applying another Hadamard gate on the bottom ancillary qubit yield

\begin{equation}
\ket{\psi_{im}}=|i\rangle|m\rangle \left((\ket{\boldsymbol{z'}_{i}}\ket{0}_{\textsf{other}}+...)+\ket{\boldsymbol{w}_{m}}\ket{0}_{\textsf{other}}\right)\ket{0}+|i\rangle|m\rangle \left((\ket{\boldsymbol{z'}_{i}}\ket{0}_{\textsf{other}}+...)-\ket{\boldsymbol{w}_{m}}\ket{0}_{\textsf{other}}\right)\ket{1}
\end{equation}

Denote $|u_{im}\rangle$ and $|v_{im}\rangle$ as the normalized states of $\left((\ket{\boldsymbol{z'}_{i}}\ket{0}_{\textsf{other}}+...)+\ket{\boldsymbol{w}_{m}}\ket{0}_{\textsf{other}}\right)$ and \\ $\left((\ket{\boldsymbol{z'}_{i}}\ket{0}_{\textsf{other}}+...)-\ket{\boldsymbol{w}_{m}}\ket{0}_{\textsf{other}}\right)$ respectively.\newline

Then there is a real number $\theta_{im}$ such that
\begin{align}
\ket{\psi_{im}}=|i\rangle|m\rangle (\underbrace{\left.\left.\sin \theta_{im}\left|u_{im}\right\rangle\ket{0}+\cos \theta_{im}\left|v_{im}\right\rangle|1\right\rangle\right)}_{\mid \phi_{im}\rangle} =|i\rangle|m\rangle  \left|\phi_{im}\right\rangle
\end{align}
 $\theta_{im}$ satisfies $\cos\theta_{im}=\sqrt{1- \langle\boldsymbol{z}'_{i} |\boldsymbol{w}_{m}\rangle}/\sqrt{2}$, $\sin\theta_{im}=\sqrt{1+ \langle\boldsymbol{z}'_{i} |\boldsymbol{w}_{m}\rangle}/\sqrt{2}$, and we have:
\begin{equation}
\langle\boldsymbol{z}'_{i} |\boldsymbol{w}_{m}\rangle=-\cos{2\theta_{im}}.
\end{equation}
To summarize, the quantum circuit depicted in Fig.\ref{feed1}, denoted as $U$, acts as

\begin{equation}
    U: |i\rangle|m\rangle \ket{0}_{k,h}   \ket{0}_{\textsf{other}}\ket{0}\to |i\rangle|m\rangle (\underbrace{\left.\left.\sin \theta_{im}\left|u_{im}\right\rangle\ket{0}+\cos \theta_{im}\left|v_{im}\right\rangle|1\right\rangle\right)}_{\mid \phi_{im}\rangle} =|i\rangle|m\rangle  \left|\phi_{im}\right\rangle, 
    \label{actionU}
\end{equation}

where $\boldsymbol{y}_{i}^{(m)}=\bra{\boldsymbol{z}'_{i} }\boldsymbol{w}_{m}\rangle$  are encoded as
\begin{equation}\label{relationi}
\langle\boldsymbol{z}'_{i} |\boldsymbol{w}_{m}\rangle=-\cos{2\theta}_{im}.
\end{equation}

\begin{figure}[h!]
    \centering
    \includegraphics[width=\linewidth]{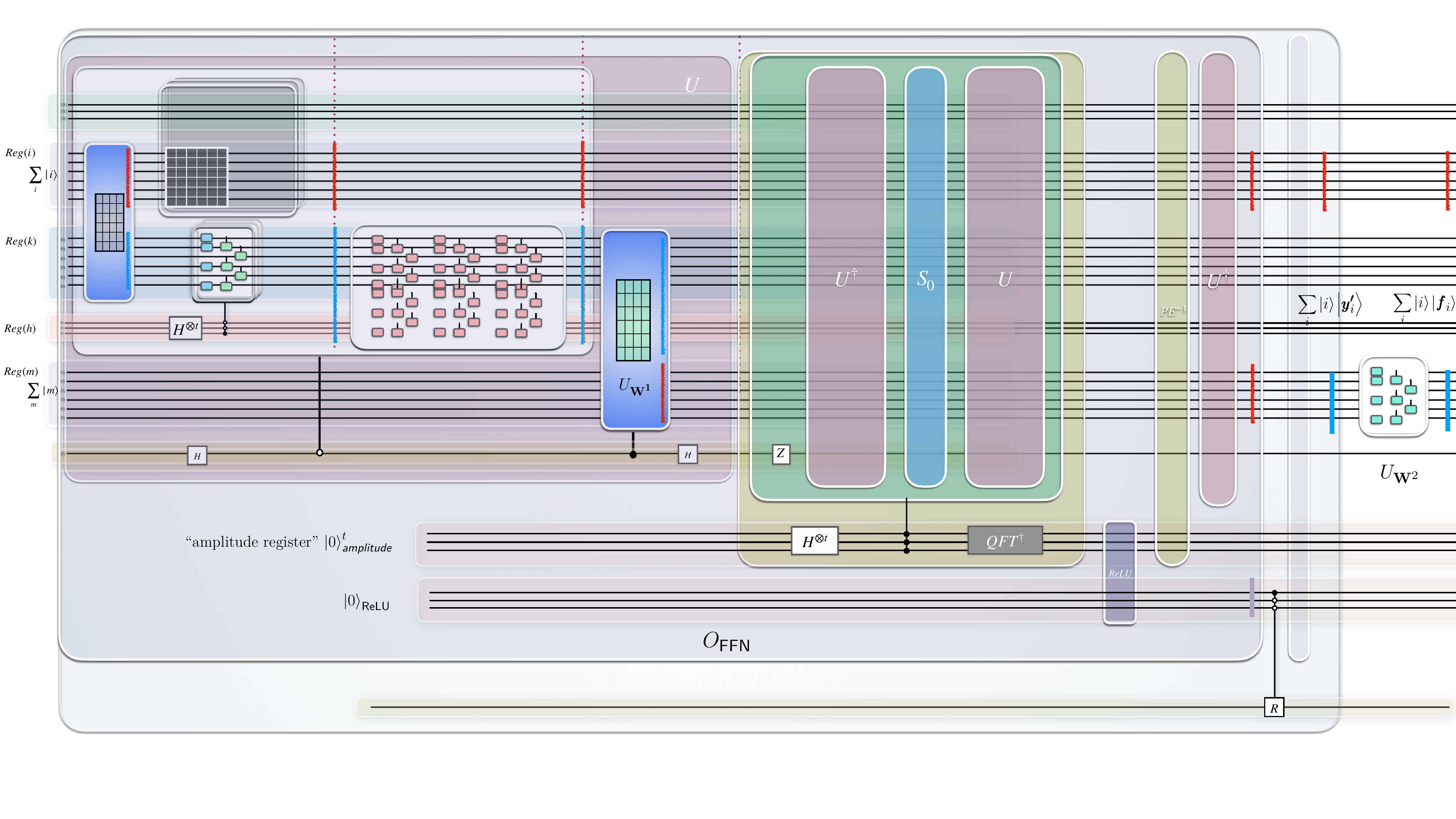}
\caption{\textit{Feed-Forward Network on Quantum computer 2} The figure illustrates the description from Eqn.\ref{actionU} to \ref{final}.The pink box in this figure, denoted as $U$, is meant to be the circuit in Fig.\ref{feed1}, but for simplicity, we omitted the residual connection as in Fig.\ref{feed1}. However, the derivation follows the same logic.  }
    \label{feed2}
\end{figure}

When acting on 
the input state to the circuit $\left|\Psi_{0}\right\rangle=\sum_{i}\sum_{m}|i\rangle|m\rangle \ket{0}_{k,h}   \ket{0}_{\textsf{other}}\ket{0}$, $U$ , also depicted as the pink box in Fig.\ref{feed2}, produces the following state

\begin{equation}
\left|\Psi_{1}\right\rangle=\sum_{i}\sum_{m}|i\rangle|m\rangle (\underbrace{\left.\left.\sin \theta_{im}\left|u_{im}\right\rangle\ket{0}+\cos \theta_{im}\left|v_{im}\right\rangle|1\right\rangle\right)}_{\mid \phi_{im}\rangle} = \sum_{i}\sum_{m}|i\rangle|m\rangle  \left|\phi_{im}\right\rangle
\label{amplitudeencoding1}
\end{equation}

Next use amplitude estimation \cite{brassard2002quantum} to extract and store $\boldsymbol{y}_{i}^{(m)}=\bra{\boldsymbol{z}'_{i} }\boldsymbol{w}_{m}\rangle$ into an additional register which we call the ``amplitude register'' $\ket{0}^{t}_{\textsf{\textit{amplitude}}}$ and
the output state $\ket{\Psi_{1}}$ (using the same notation) becomes
\begin{align}
  \left|\Psi_{3}\right\rangle=\sum_{i}\sum_{m}|i\rangle|m\rangle  \left|\phi_{im}\right\rangle\ket{0}^{t}_{\textsf{\textit{amplitude}}},
\end{align}
where $\ket{\phi_{im}}$ can be decomposed as
\begin{equation}
\ket{\phi_{im}}=\frac{-i}{\sqrt{2}}\left(e^{i \theta_{im}}\ket{\omega_{+}}_{im}-e^{i(-\theta_{im})}\ket{\omega_{-}}_{im}\right).
\end{equation}

where
$
|w_{\pm}\rangle_{im}=\frac{1}{\sqrt{2}}(|0\rangle|u_{im}\rangle\pm\bm i|1\rangle|v_{im}\rangle).
$\newline

Hence, we have
\begin{equation}
\left|\Psi_{1}\right\rangle=\sum_{i}\sum_{m}\frac{-i}{\sqrt{2}}\left( e^{i \theta_{im}}|i\rangle\left|m\right\rangle\ket{\omega_{+}}_{im}-e^{i(-\theta_{im})}|i\rangle\left|m\right\rangle\ket{\omega_{-}}_{im}\right) \ket{0}^{t}_{\textsf{\textit{amplitude}}}.
\label{phi1}
\end{equation}

The overall Grover operator $G$ is defined as
\begin{equation}
G \coloneqq UC_{2}U^{-1}C_{1},
\label{grovero2}
\end{equation}
where $C_{1}$ is the $Z$ gate on the bottom ancilla qubit in the pink box, and $C_{2}=(I-2|0\rangle\langle0|)\otimes I_{i,m}$ is the ``flip zero state'' on registers other than $Reg(i), Reg(m)$ (represented as $S_0$ in Fig.\ref{feed2}).\newline

Utilising Eqn. \ref{actionU}, it can be shown that $G$ can be expressed as
\begin{align}
    G=\sum_{i}\sum_{m}\ket{i}\left|m \left>\right<m\right|\bra{i}\otimes G_{im},
\end{align}
where $G_{im}$ is defined as

\begin{equation}
    G_{im}= (I-2|\phi_{im}\rangle\langle\phi_{im}|))(Z\otimes I)
\end{equation}

It is easy to check that $|w_{\pm}\rangle_{im}$ are the eigenstates of $G_{im}$, that is,
\begin{align}\label{eq_G22}
G_{im}|w_{\pm}\rangle_{im}=e^{\pm\bm i  2\theta_{im}}|w_{\pm}\rangle_{im}.
\end{align}
Therefore, the overall Grover operator $G$ possesses the following eigen-relation:
\begin{align}
 G\ket{i}\left|m\right\rangle\ket{\omega_{\pm}}_{im}=   e^{i (\pm 2\theta_{im})}\ket{i}\left|m\right\rangle\ket{\omega_{\pm}}_{im}.
\end{align}

Next, we apply phase estimation of the overall Grover operator $G$ on the input state
$\left|\Psi_{1}\right\rangle$. The resulting state $\left|\Psi_{2}\right\rangle$ can be written as
\begin{equation}\label{statephase}
\left|\Psi_{2}\right\rangle=\sum_{i}\sum_{m}\frac{-i}{\sqrt{2}}\left( e^{i \theta_{im}}\ket{i}\left|m\right\rangle\ket{\omega_{+}}_{im}\ket{2\theta_{im}}-e^{i(-\theta_{im})}\ket{i}\left|m\right\rangle\ket{\omega_{-}}_{im}\ket{-2\theta_{im}}\right).
\end{equation}

Note here in Eq.~\ref{statephase}, $\ket{\pm 2\theta_{im}}$ denotes the eigenvalues $\pm 2\theta_{im}$ being stored in the amplitude register with some finite precision.\newline

Then we apply an oracle $U_{O}$ (implemented by arithmetic circuit) on the amplitude register and an extra ancilla register $\ket{0}_{\textsf{ReLU}}$, which acts as
\begin{equation}
 U_{O}\ket{0}_{\textsf{ReLU}}\ket{\pm 2\theta_{im}}=\ket{\operatorname{ReLU}(\boldsymbol{y}_{i}^{(m)})}\ket{\pm2\theta_{im}},
 \label{relu}
\end{equation}

The state after the oracle can be written as

\begin{equation}
\left|\Psi_{3}\right\rangle=\sum_{i}\sum_{m}\frac{-i}{\sqrt{2}}\ket{\operatorname{ReLU}(\boldsymbol{y}_{i}^{(m)})}\left( e^{i \theta_{im}}\ket{i}\left|m\right\rangle\ket{\omega_{+}}_{im}\ket{2\theta_{im}}-e^{i(-\theta_{im})}\ket{i}\left|m\right\rangle\ket{\omega_{-}}_{im}\ket{-2\theta_{im}}\right).
\end{equation}

Then we perform the uncomputation of Phase estimation, resulting in the state,
\begin{align}
\left|\Psi_{4}\right\rangle=\sum_{i}\sum_{m}\frac{-i}{\sqrt{2}}\ket{\operatorname{ReLU}(\boldsymbol{y}_{i}^{(m)})}\left( e^{i \theta_{im}}\ket{i}\left|m\right\rangle\ket{\omega_{+}}_{im}\ket{0}^{t}_{\textsf{\textit{amplitude}}}-e^{i(-\theta_{im})}\ket{i}\left|m\right\rangle\ket{\omega_{-}}_{im}\ket{0}^{t}_{\textsf{\textit{amplitude}}}\right)\\ =\sum_{i}\sum_{m}\ket{\operatorname{ReLU}(\boldsymbol{y}_{i}^{(m)})}|i\rangle|m\rangle  \left|\phi_{im}\right\rangle\ket{0}^{t}_{\textsf{\textit{amplitude}}}
\end{align}

Finally, we perform $U^{\dag}$ and the resulting state is
\begin{equation}
\left|\Psi_{5}\right\rangle=\sum_{i}\sum_{m}\ket{\operatorname{ReLU}(\boldsymbol{y}_{i}^{(m)})}|i\rangle|m\rangle  \ket{0}_{k,h}   \ket{0}_{\textsf{other}}\ket{0}\ket{0}^{t}_{\textsf{\textit{amplitude}}}.
\label{dec}
\end{equation}

The above steps, as gathered in the grey box in Fig.~\ref{feed2}, implement an oracle $O_{\textsf{FFN}}$ such that:
\begin{equation}
  O_{\textsf{FFN}}  : \ket{i}\ket{m}\ket{0}_{k,h}   \ket{0}_{\textsf{other}}\ket{0}\ket{0}_{\textsf{ReLU}} \to \ket{i}\ket{m}\ket{0}_{k,h}   \ket{0}_{\textsf{other}}\ket{0}\ket{\operatorname{ReLU}(\boldsymbol{y}_{i}^{(m)})}
\end{equation}

Which produces the state

\begin{equation}
\sum_i\sum_m \ket{i}\ket{m}\ket{\boldsymbol{y'}_{i}^{(m)}}\ket{0}_{k,h}   \ket{0}_{\textsf{other}}\ket{0}
\end{equation}

where we denote $\boldsymbol{y'}_{i}^{(m) }=\operatorname{ReLU}(\boldsymbol{y}_{i}^{(m)})$. \newline

Next, the "Conditional Rotation" (Theorem 3.5 in Ref.~\cite{landman2021quantum}) and uncomputation of $O_{\textsf{FFN}}$ are applied, obtaining

\begin{equation}
\sum_i\sum_m \ket{i}\boldsymbol{y'}_{i}^{(m)}\ket{m}=\sum_i \ket{i}\sum_m\boldsymbol{y'}_{i}^{(m)}\ket{m}=\sum_i \ket{i}\ket{\boldsymbol{y'}_{i}}\end{equation}

where $\ket{\boldsymbol{y'}_{i}}:=\sum_m\boldsymbol{y'}_{i}^{(m)}\ket{m}$ and we omitted the registers being in "all-zero" state. \newline

Finally, a trainable unitary $U_{\mathbf{W}^2}$ (a parameterised quantum circuit) implementing ${\mathbf{W}^2}^{\top}$ is applied

\begin{equation}
\left|\Psi_{\boldsymbol{F}}\right\rangle=\sum_i \ket{i}U_{\mathbf{W}^2}\ket{\boldsymbol{y'}_{i}}=\sum_i \ket{i}\ket{\boldsymbol{f}_{i}}
\label{final}
\end{equation}

where $\boldsymbol{f}_{i}:={\mathbf{W}^2}^{\top}\boldsymbol{y'}_{i}=\operatorname{FFN}\left(\boldsymbol{z}'_{i}\right) \in \mathbb{R}^{rH}$ and we define $\boldsymbol{F}:=\left[\boldsymbol{f}_{1}, \ldots, \boldsymbol{f}_{n}\right] \in$ $\mathbb{R}^{rH \times n}$. \newline

By the end of the quantum circuit in Fig.~\ref{feed2}, we have obtained the final state from a Transformer block.\newline

A tomography procedure (Theorem 4.3 from Ref.\cite{landman2021quantum}) is performed to read out the amplitudes of the final state of a Transformer block. The results are then used as the input for the next Transformer block.

\subsection{Generative Pre-training on Quantum Computer}\label{generative}

Using the same notation as in the previous section, we denote the final state from the last Transformer block in GPT as:

\begin{equation}
\left|\Psi_{\boldsymbol{F}}\right\rangle=\sum_i \ket{i}\ket{\boldsymbol{f}_{i}}
\end{equation}

where $\boldsymbol{f}_{i}\in \mathbb{R}^{rH}$ and $\boldsymbol{F}:=\left[\boldsymbol{f}_{1}, \ldots, \boldsymbol{f}_{n}\right] \in$ $\mathbb{R}^{rH \times n}$.\newline

For language modeling, we then apply a linear layer $\boldsymbol{W}_E\in \mathbb{R}^{V \times rH}$ to map the output back to the vocabulary space as $\boldsymbol{f'}_{i}=\boldsymbol{W}_E \boldsymbol{f}_{i} \in \mathbb{R}^{V}$.  $\boldsymbol{W}_E$ can be implemented on a quantum circuit by applying a trainable unitary $U_{\boldsymbol{W}_E}$ (a parameterised quantum circuit) on $Reg(k), Reg(m)$ and some extra qubits (since $V \gg rH$) and we obtain the state 

\begin{equation}
\left|\Psi_{\boldsymbol{F'}}\right\rangle=U_{\boldsymbol{W}_E}(\left|\Psi_{\boldsymbol{F}}\right\rangle \otimes \ket{0}) =\sum_i \ket{i}U_{\boldsymbol{W}_E}(\ket{\boldsymbol{f}_{i}}\otimes \ket{0})=\sum_i \ket{i}\ket{\boldsymbol{f'}_{i}}		
\end{equation}

where $\boldsymbol{F'}:=\left[\boldsymbol{f'}_{1}, \ldots, \boldsymbol{f'}_{n}\right] \in$ $\mathbb{R}^{V \times n}$.\newline

For the $t$th batch in Generative Pre-training, we examine $\boldsymbol{f'}_{t+1}$ in the output, given the input $\left[\boldsymbol{x}_{1}, \ldots, \boldsymbol{x}_{t}\right]$. The loss function for this batch is defined as the cross-entropy between $\operatorname{softmax}(\boldsymbol{f'}_{t+1})$ and the one-hot encoding of the $(t+1)$th token in the training text, denoted by $\boldsymbol{b}_{t+1} \in \mathbb{B}^{V}$. On the quantum circuit, the loss function can be correspondingly defined as the overlap between $\ket{\boldsymbol{f'}_{t+1}}$ and $\ket{\boldsymbol{b}_{t+1}}$\footnote{$\ket{\boldsymbol{f'}_{t+1}}$ and $\ket{\boldsymbol{b}_{t+1}}$ represent the amplitude encoding of $\boldsymbol{f'}_{t+1}$ and $\boldsymbol{b}_{t+1}$.}, which can be evaluated via swap test on $\left|\Psi_{\boldsymbol{F'}}\right\rangle = \sum_i \ket{i}\ket{\boldsymbol{f'}_{i}}$ and an additional state $\left|\Psi_{\boldsymbol{b}_{t+1}}\right\rangle = \ket{t+1}\ket{\boldsymbol{b}_{t+1}}$. The cumulative loss is calculated across all batches, followed by the execution of certain optimization methods to adjust the model’s parameters.

\section{Conclusion}\label{conclusion}

In this paper, we have presented a comprehensive framework for translating key components of the GPT architecture---namely, the masked multi-head attention module, the feed-forward networks module, and the residual connection---along with the generative pre-training phase, into the quantum computing paradigm. This translation demonstrates the feasibility of implementing complex language models on quantum computers, marking an initial step toward harnessing the power of quantum computing to enhance the capabilities of LLMs. As we progress, it becomes imperative to conduct a detailed analysis of the quantum resources required for our quantum implementation of GPT, such as the number of qubits and circuit depths, and to perform a thorough comparison with classical GPT implementations. \newline

In summary, the integration of quantum computing with LLMs, as demonstrated by our work, opens new avenues for research and development in the field of QML. By continuing to explore these possibilities, we move closer to realizing the full potential of quantum-enhanced artificial intelligence, setting the stage for future breakthroughs in computational efficiency and model capabilities.\newline

\bibliography{Ref.bib}

\appendix
\section{Quantum Attention Oracle }\label{attentionsec}

In this section, we introduce the construction of the quantum attention oracle $O_{\textsf{attention}}$, as mentioned in Section~\ref{selfsection}. The quantum attention oracle $O_{\textsf{attention}}$ is designed to coherently evaluate and store the attention score $a(\bold{x}_i,\bold{x}_j)$ for each pair of input vectors $\bold{x}_i$ and $\bold{x}_j$. It acts as follows:
\begin{equation}
  O_{\textsf{attention}}  \ket{i}\ket{j}\ket{0} \to \ket{i}\ket{j}\ket{a(\bold{x}_i,\bold{x}_j)}.
  \label{attentionoracle}
\end{equation}

The construction of $O_{\textsf{attention}}$ consists of the following two steps:

\subsection{Evaluating Attention score in superposition}\label{step1}

The attention score $a(\bold{x}_i,\bold{x}_j)$ can take one of the standard forms in the classical literature \cite{ghojogh2020attention} --- the inner product of the linearly transformed input vectors:

\begin{equation}
	a(\bold{x}_i,\bold{x}_j)=\bold{x}_i^T W_K^T W_Q \bold{x}_j, 
 \label{attention-score }
	\end{equation}
where $W_K, W_Q$ are trainable linear transformations. \\

In terms of Dirac notation, this can be written as:

\begin{equation}
	a(\bold{x}_i,\bold{x}_j)=\bra{\bold{x}_i}U_K^{\dagger} U_Q \ket{\bold{x}_j}\end{equation}

in which $U_K, U_Q$ are trainable unitaries, $\ket{\bold{x}_{i}}:=\sum_{k=1}^{d}\bold{x}^{(k)}_{i}|k\rangle$ is the amplitude encoding of the vector $\bold{x}_{i}$ whose $k$-th elements are denoted as $\bold{x}^{(k)}_{i}$. Note here and throughout the paper, we omit the normalization factors.\newline

\begin{figure}[h!]
    \centering
    \includegraphics[width=\linewidth]{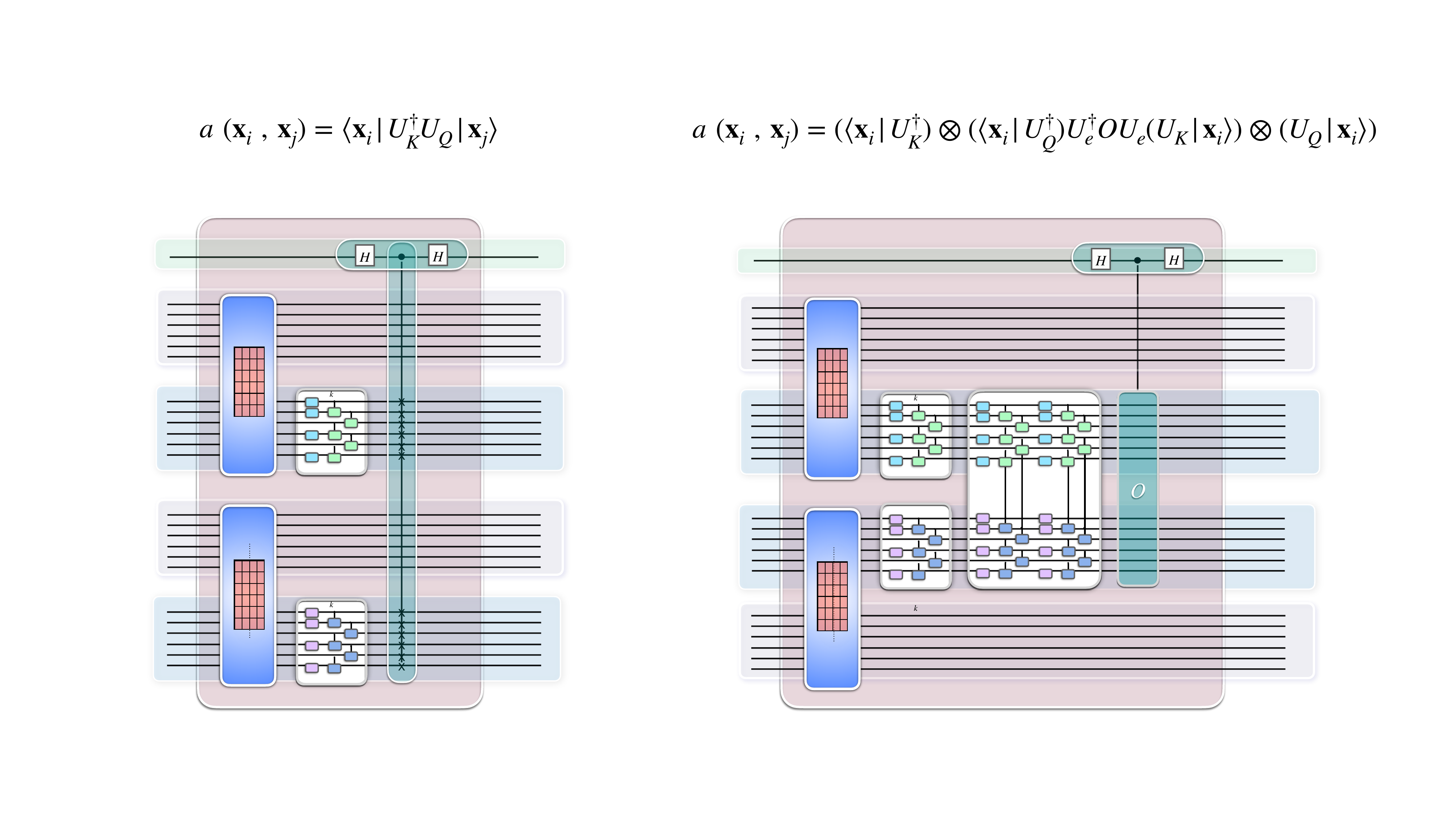}
   \caption{\textit{Evaluating Attention score in superposition} The attention score $a(\bold{x}_i,\bold{x}_j)$ can take one of the standard forms in the classical literature \cite{ghojogh2020attention} --- the inner product of the linearly transformed input vectors:  $a(\bold{x}_i,\bold{x}_j)=\bold{x}_i^T W_K^T W_Q \bold{x}_j$, in which $W_K, W_Q$ are trainable linear transformations. This attention score can be evaluated in superposition on the quantum circuit by parallel swap test \cite{liao2021quantum}, depicted as the left side of this figure. On the right side of this figure, we illustrate an alternative form of the Attention score, which can be evaluated by parallel Hadamard test \cite{liao2021quantum}. }
    \label{att}
\end{figure}

This attention score can be evaluated on a quantum circuit by parallel swap test \cite{liao2021quantum} as depicted in Fig.~\ref{att} which we will discuss in detail below.\\

We denote the unitary for the parallel swap test circuit, as circled by the pink box on the left side of Fig.~\ref{att}, as $U$.
The input to $U$, $\ket{\Psi_{0}}$, can be written as (note here and throughout the paper, we omit the normalization factors):

\begin{equation}
\left|\Psi_{0}\right\rangle=\ket{0}\otimes(\sum_{i}\left|i\right\rangle)\otimes \left|0\right\rangle^{n}_\mathtt{K} \otimes(\sum_{j}\left|j\right\rangle)\otimes\left|0\right\rangle^{n}_\mathtt{Q} 
\end{equation}

where $\left|0\right\rangle^{n}_\mathtt{K}, \left|0\right\rangle^{n}_\mathtt{Q}$ are the initial states of the two registers on which the input vectors will be loaded. The input encoding via Controlled Quantum State Preparation\cite{yuan2023optimal}(mentioned in Section~\ref{selfsection}), depicted as the blue boxes in Fig.\ref{att}, can be written as $\sum_{i}\left|i\right\rangle \bra{i}\otimes U_{\bold{x}_i}$ where $U_{\bold{x}_i}$ act as $U_{\bold{x}_i}\ket{0}=\left|\bold{x}_i\right\rangle$.\newline

Applying this input encoding results in the following state:

\begin{equation}
\left|\Psi_{1}\right\rangle=\ket{0}\otimes(\sum_{i}\left|i\right\rangle\otimes \left|\bold{x}_i\right\rangle^{n} )\otimes(\sum_{j}\left|j\right\rangle\otimes\left|\bold{x}_j\right\rangle^{n} )
\end{equation}

Then, $U_K, U_Q$ implemented by parametrized quantum circuits are applied on the two registers hosting the input vectors, yielding the following state:

\begin{equation}
\left|\Psi_{2}\right\rangle=\ket{0}\otimes(\sum_{i}\left|i\right\rangle\otimes U_K \left|\mathbf{x}_i\right\rangle^{n} )\otimes(\sum_{j}\left|j\right\rangle\otimes U_Q\left|\bold{x}_j\right\rangle^{n} )
\end{equation}

We further define $\mathcal{K}_i,\mathcal{Q}_j$ and corresponding state $\ket{k_i}, \ket{q_j} $ as
$\mathcal{K}_i\ket{0}_\mathtt{K}^{n}=U_K \left|\mathbf{x}_i\right\rangle=\ket{k_i},\mathcal{Q}_j\ket{0}_\mathtt{Q}^{n}=U_Q\left|\mathbf{x}_j\right\rangle=\ket{q_j}$. Then $U$ can be written explicitly as

\begin{multline}
U\coloneqq[H\otimes I\otimes I\otimes I\otimes I]\cdot\\
[\ket{0}\bra{0}\otimes (\sum_{i}\sum_{j}\left|i \left>\right<i\right|\otimes \mathcal{K}_{i}\otimes\left|j \left>\right<j\right|\otimes \mathcal{Q}_{j} )+\ket{1}\bra{1}\otimes (\sum_{i}\sum_{j}\left|i \left>\right<i\right|\otimes \mathcal{Q}_{j}\otimes\left|j \left>\right<j\right|\otimes  \mathcal{K}_{i} )]\\\cdot[H\otimes I\otimes I\otimes I\otimes I].
\label{defineu}
\end{multline}

Reorganizing the terms in Eqn~\ref{defineu} we have

\begin{align}
 U=\sum_{i}\sum_j\left|i \left>\right<i\right|\otimes \left|j \left>\right<j\right|\otimes U_{ij},
\end{align}

where

\begin{equation}
U_{ij}\coloneqq[H\otimes I\otimes I]\cdot
[\ket{0}\bra{0}\otimes  \mathcal{K}_{i}\otimes \mathcal{Q}_{j} +\ket{1}\bra{1}\otimes  \mathcal{Q}_{j}\otimes  \mathcal{K}_{i}\ ]\cdot[H\otimes  I\otimes I],
\end{equation}
\\

Define $\ket{\phi_{ij}} \coloneqq U_{ij}\ket{0}\left|0\right\rangle^{n}_{\mathtt{K}} \left|0\right\rangle^{n}_{\mathtt{Q}}$ and we have:
\begin{equation}
  |\phi_{ij}\rangle=\frac{1}{\sqrt{2}}(\ket{+}\ket{k_i}\ket{q_j}+\ket{-}\ket{q_j}\ket{k_i}).
  \label{phiij}
\end{equation}

Expanding and rearranging the terms in 
Eq.~\ref{phiij} we have
\begin{equation}
  |\phi_{ij}\rangle=\frac{1}{2}\left(|0\rangle\otimes(|k_i\rangle\ket{q_j}+|q_j\rangle\ket{k_i})+|1\rangle\otimes(|k_i\rangle\ket{q_j}-|q_j\rangle\ket{k_i}\right).
\label{sw}
\end{equation}

Denote $|u_{ij}\rangle$ and $|v_{ij}\rangle$ as the normalized states of $|k_i\rangle\ket{q_j}+|q_j\rangle\ket{k_i}$ and $|k_i\rangle\ket{q_j}-|q_j\rangle\ket{k_i}$ respectively.
Then there is a real number $\theta_{ij}\in[{\pi}/{4},{\pi}/{2}]$ such that
\begin{align}\label{amplitudeencoding}
|\phi_{ij}\rangle=\sin\theta_{ij}|0\rangle|u_{ij}\rangle+\cos\theta_{ij}|1\rangle|v_{ij}\rangle.
\end{align}
 $\theta_{ij}$ satisfies $\cos\theta_{ij}=\sqrt{1- |\langle k_i|q_j\rangle|^2}/\sqrt{2}$, $\sin\theta_{ij}=\sqrt{1+ |\langle k_i|q_j\rangle|^2}/\sqrt{2}$.\\

The final output state from $U$,  $\left|\Psi_{3}\right\rangle=U\left|\Psi_{0}\right\rangle $, can then be written as

\begin{equation}
\left|\Psi_{3}\right\rangle=\sum_{i}\sum_{j}|i\rangle|j\rangle (\underbrace{\left.\left.\sin \theta_{ij}\left|u_{ij}\right\rangle\ket{0}+\cos \theta_{ij}\left|v_{ij}\right\rangle|1\right\rangle\right)}_{\mid \phi_{ij}\rangle} = \sum_{i}\sum_{j}|i\rangle|j\rangle  \left|\phi_{ij}\right\rangle
\label{amplitudeencoding2}
\end{equation}

Note that $\langle k_i|q_j\rangle=\bra{\bold{x}_i}U_K^{\dagger} U_Q \ket{\bold{x}_j}=a(\bold{x}_i,\bold{x}_j)$ being the attention scores are encoded in the amplitudes of the output state $\left|\Psi_{3}\right\rangle$ of swap test as:
\begin{equation}
 |\langle k_i|q_j\rangle|^2=-\cos{2\theta_{ij}}.
\label{relation}
\end{equation}

\subsection{Storing Attention score}\label{step2}

The second step is to use amplitude estimation \cite{brassard2002quantum} to extract and store the attention scores into an additional register which we call the ``amplitude register''. \newline

After step 1, we introduce an extra register $\ket{0}^{t}_{\textsf{\textit{amplitude}}}$ and
the output state $\ket{\Psi_{3}}$ (using the same notation) becomes
\begin{align}
  \left|\Psi_{3}\right\rangle=\sum_{i}\sum_{j}|i\rangle|j\rangle  \left|\phi_{ij}\right\rangle\ket{0}^{t}_{\textsf{\textit{amplitude}}},
\end{align}
where $\ket{\phi_{ij}}$ can be decomposed as
\begin{equation}
\ket{\phi_{ij}}=\frac{-i}{\sqrt{2}}\left(e^{i \theta_{ij}}\ket{\omega_{+}}_{ij}-e^{i(-\theta_{ij})}\ket{\omega_{-}}_{ij}\right).
\end{equation}
Hence, we have
\begin{equation}
\left|\Psi_{3}\right\rangle=\sum_{i}\sum_{j}\frac{-i}{\sqrt{2}}\left( e^{i \theta_{ij}}|i\rangle\left|j\right\rangle\ket{\omega_{+}}_{ij}-e^{i(-\theta_{ij})}|i\rangle\left|j\right\rangle\ket{\omega_{-}}_{ij}\right) \ket{0}^{t}_{\textsf{\textit{amplitude}}}.
\label{phi12}
\end{equation}

\begin{figure}[h!]
    \centering
    \includegraphics[width=0.9\linewidth]{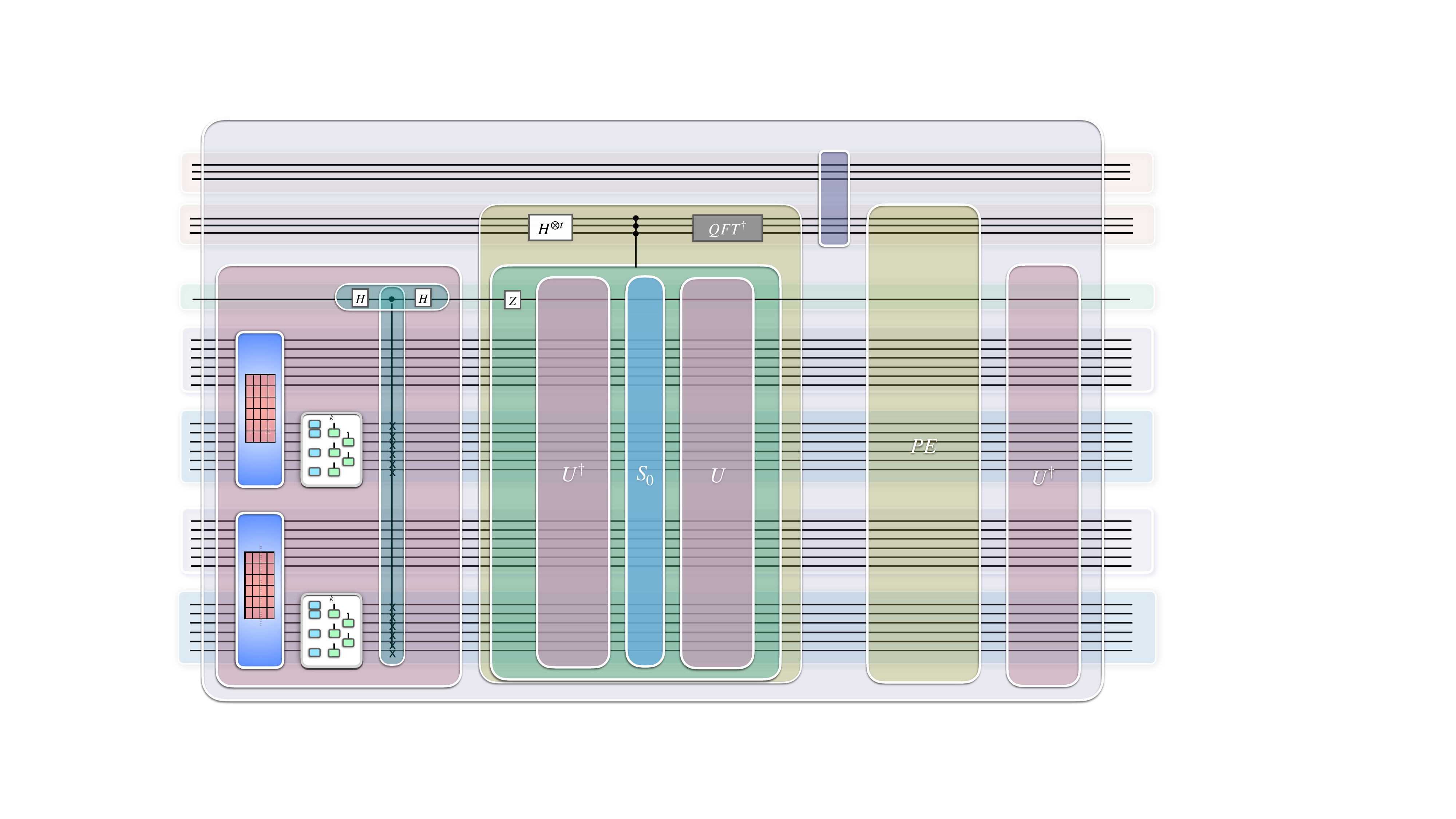}
   \caption{\textit{Quantum  attention oracle $O_{\textsf{attention}}$} The quantum attention oracle aims to coherently evaluate and store attention score $a(\bold{x}_i,\bold{x}_j)$ for each pair of the input vectors, it acts as $
  O_{\textsf{attention}}  \ket{i}\ket{j}\ket{0} \to \ket{i}\ket{j}\ket{a(\bold{x}_i,\bold{x}_j)}
$. The construction of the quantum attention oracle, depicted in this figure, is detailed in the appendix subsections \ref{step1} and \ref{step2}.}
    \label{att23}
\end{figure}

The overall Grover operator $G$ is defined as
\begin{equation}
G \coloneqq UC_{2}U^{\dagger}C_{1},
\label{grovero}
\end{equation}
where $C_{1}$ is the $Z$ gate on the swap ancilla qubit, and $C_{2}=I-2|0\rangle\langle0|$ is the ``flip zero state'' on registers other than the two registers hosting indices $i,j$ (represented as $S_0$ in Fig.\ref{att23}).
It can be shown that $G$ can be expressed as
\begin{align}
    G=\sum_{i}\sum_{j}\ket{i}\left|j \left>\right<j\right|\bra{i}\otimes G_{ij},
\end{align}
where $G_{ij}$ is defined as

\begin{equation}
    G_{ij}= (I-2|\phi_{ij}\rangle\langle\phi_{ij}|))C_{1}\end{equation}

It is easy to check that $|w_{\pm}\rangle_{ij}$ are the eigenstates of $G_{ij}$, that is,
\begin{align}\label{eq_G2}
G_{ij}|w_{\pm}\rangle_{ij}=e^{\pm\bm i  2\theta_{ij}}|w_{\pm}\rangle_{ij}.
\end{align}
The overall Grover operator $G$ possesses the following eigen-relation:
\begin{align}
 G\ket{i}\left|j\right\rangle\ket{\omega_{\pm}}_{ij}=   e^{i (\pm 2\theta_{ij})}\ket{i}\left|j\right\rangle\ket{\omega_{\pm}}_{ij}.
\end{align}

Next, we apply phase estimation of the overall Grover operator $G$ on the input state
$\left|\Psi_{3}\right\rangle$. The resulting state $\left|\Psi_{4}\right\rangle$ can be written as
\begin{equation}\label{statephase2}
\left|\Psi_{4}\right\rangle=\sum_{i}\sum_{j}\frac{-i}{\sqrt{2}}\left( e^{i \theta_{ij}}\ket{i}\left|j\right\rangle\ket{\omega_{+}}_{ij}\ket{2\theta_{ij}}-e^{i(-\theta_{ij})}\ket{i}\left|j\right\rangle\ket{\omega_{-}}_{ij}\ket{-2\theta_{ij}}\right).
\end{equation}

Note here in Eq.~\ref{statephase2}, $\ket{\pm 2\theta_{ij}}$ denotes the eigenvalues $\pm 2\theta_{ij}$ being stored in the amplitude register with some finite precision.\newline

Next, we apply an oracle $U_{O}$ (implemented by arithmetic circuit) on the amplitude register and an extra ancilla register, which acts as
\begin{equation}
 U_{O}\ket{0}\ket{\pm 2\theta_{ij}}=\ket{a(\bold{x}_i,\bold{x}_j)}\ket{\pm2\theta_{ij}},
 \label{threshold}
\end{equation}

The state after the oracle can be written as

\begin{equation}
\left|\Psi_{5}\right\rangle=\sum_{i}\sum_{j}\frac{-i}{\sqrt{2}}\ket{a(\bold{x}_i,\bold{x}_j)}\left( e^{i \theta_{ij}}\ket{i}\left|j\right\rangle\ket{\omega_{+}}_{ij}\ket{2\theta_{ij}}-e^{i(-\theta_{ij})}\ket{i}\left|j\right\rangle\ket{\omega_{-}}_{ij}\ket{-2\theta_{ij}}\right).
\end{equation}

Then we perform the uncomputation of Phase estimation, the resulting state is
\begin{align}
\left|\Psi_{6}\right\rangle=\sum_{i}\sum_{j}\frac{-i}{\sqrt{2}}\ket{a(\bold{x}_i,\bold{x}_j)}\left( e^{i \theta_{ij}}\ket{i}\left|j\right\rangle\ket{\omega_{+}}_{ij}\ket{0}^{t}_{\textsf{\textit{amplitude}}}-e^{i(-\theta_{ij})}\ket{i}\left|j\right\rangle\ket{\omega_{-}}_{ij}\ket{0}^{t}_{\textsf{\textit{amplitude}}}\right)\\ =\sum_{i}\sum_{j}\ket{a(\bold{x}_i,\bold{x}_j)}|i\rangle|j\rangle  \left|\phi_{ij}\right\rangle\ket{0}^{t}_{\textsf{\textit{amplitude}}}
\end{align}

Finally, we perform the uncomputation of the swap test and the resulting state is
\begin{equation}
\left|\Psi_{7}\right\rangle=\sum_{i}\sum_{j}\ket{a(\bold{x}_i,\bold{x}_j)}|i\rangle|j\rangle  \left|0\right\rangle\ket{0}^{t}_{\textsf{\textit{amplitude}}}.
\label{dec2}
\end{equation}

The above steps, as illustrated in Fig.~\ref{att23}, implemented the quantum attention oracle $O_{\textsf{attention}}$ such that:
\begin{equation}
  O_{\textsf{attention}}  \ket{i}\ket{j}\ket{0} \to \ket{i}\ket{j}\ket{a(\bold{x}_i,\bold{x}_j)}
\end{equation}

\section{Positional encoding via quantum circuit}\label{positional}

The positional encoding mentioned in Section \ref{tran} can be described as follows\cite{ghojogh2020attention}:
Corresponding to the $i$-th vector in the sequence $\boldsymbol{x}_{i} \in \mathbb{R}^{d}$, define the position vector $\boldsymbol{p}_{i} \in \mathbb{R}^{d}$ as:

\begin{equation}
\left\{\begin{array}{l}
\boldsymbol{p}_{i}^{(2 j+1)}:=\cos \left(\frac{i}{10000^{\frac{2 j}{d}}}\right) \\
\boldsymbol{p}_{i}^{(2 j)}:=\sin \left(\frac{i}{10000^{\frac{2 j}{d}}}\right)
\end{array}\right.
\label{definition}	
\end{equation}
for all $j \in\{0,1, \ldots,\lfloor d / 2\rfloor\}$, where $\boldsymbol{p}_{i}^{(2 j+1)}$ and $\boldsymbol{p}_{i}^{(2 j)}$ denote the odd and even elements of $\boldsymbol{p}_{i}$, respectively.  For encoding of positional information into data, the position vectors are added directly to the input vectors:

\begin{equation}
\boldsymbol{x'}_{i} = \boldsymbol{x}_{i}+\boldsymbol{p}_{i},
\label{psixp}
\end{equation}
where $\boldsymbol{x'}_{i}$ is the appended input vectors that contain positional information, and we define $\boldsymbol{X'}:=\left[\boldsymbol{x'}_{1}, \ldots, \boldsymbol{x}'_{n}\right] \in \mathbb{R}^{d \times n}$. as the matrix by stacking $\boldsymbol{x'}_{i}$.\newline

Our quantum algorithm for positional encoding aims to create the quantum state 

\begin{equation}
\left|\psi_{\boldsymbol{X'}}\right\rangle:=\sum_{i=1}^{n} |i\rangle\ket{\boldsymbol{x'}_{i}},
\label{xpr}
\end{equation}
where $\ket{\boldsymbol{x'}_{i}}:=\sum_{k=1}^{d}\boldsymbol{x'}^{(k)}_{i}|k\rangle$ is the amplitude encoding of the vector $\boldsymbol{x'}_{i}$ whose $k$-th elements are denoted as $\boldsymbol{x'}^{(k)}_{i}$. Similarly, we define $\ket{\boldsymbol{p}_{i}}:=\sum_{k=1}^{d}\boldsymbol{p}^{(k)}_{i}|k\rangle$ which is the amplitude encoding of the vector $\boldsymbol{p}_{i}$ whose $k$-th elements are denoted as $\boldsymbol{p}^{(k)}_{i}$ and 

\begin{equation}\left|\psi_{\boldsymbol{P}}\right\rangle:=\sum_{i=1}^{n} |i\rangle\ket{\boldsymbol{p}_{i}}
	\label{psip}
\end{equation}

From the above definitions of $\ket{\boldsymbol{x'}_{i}},\ket{\boldsymbol{p}_{i}}$ and Eqn.~\ref{psix},\ref{psixp},\ref{xpr},\ref{psip} we have

\begin{equation}
\left|\psi_{\boldsymbol{X'}}\right\rangle=\left|\psi_{\boldsymbol{X}}\right\rangle+\left|\psi_{\boldsymbol{P}}\right\rangle=\sum_{i=1}^{n} |i\rangle\ket{\boldsymbol{x}_{i}}+\sum_{i=1}^{n} |i\rangle\ket{\boldsymbol{p}_{i}}
\end{equation}

Note that

\begin{equation}
\ket{\boldsymbol{p}_{i}}=\sum_{k=1}^{d}\boldsymbol{p}_{i}^{(k)}|k\rangle=\sum_{j}(\boldsymbol{p}_{i}^{(2j)}|2j\rangle+\boldsymbol{p}_{i}^{(2j+1)}|2j+1\rangle)
\label{pe1}
\end{equation}

\begin{figure}[h!]
    \centering
    \includegraphics[width=\linewidth]{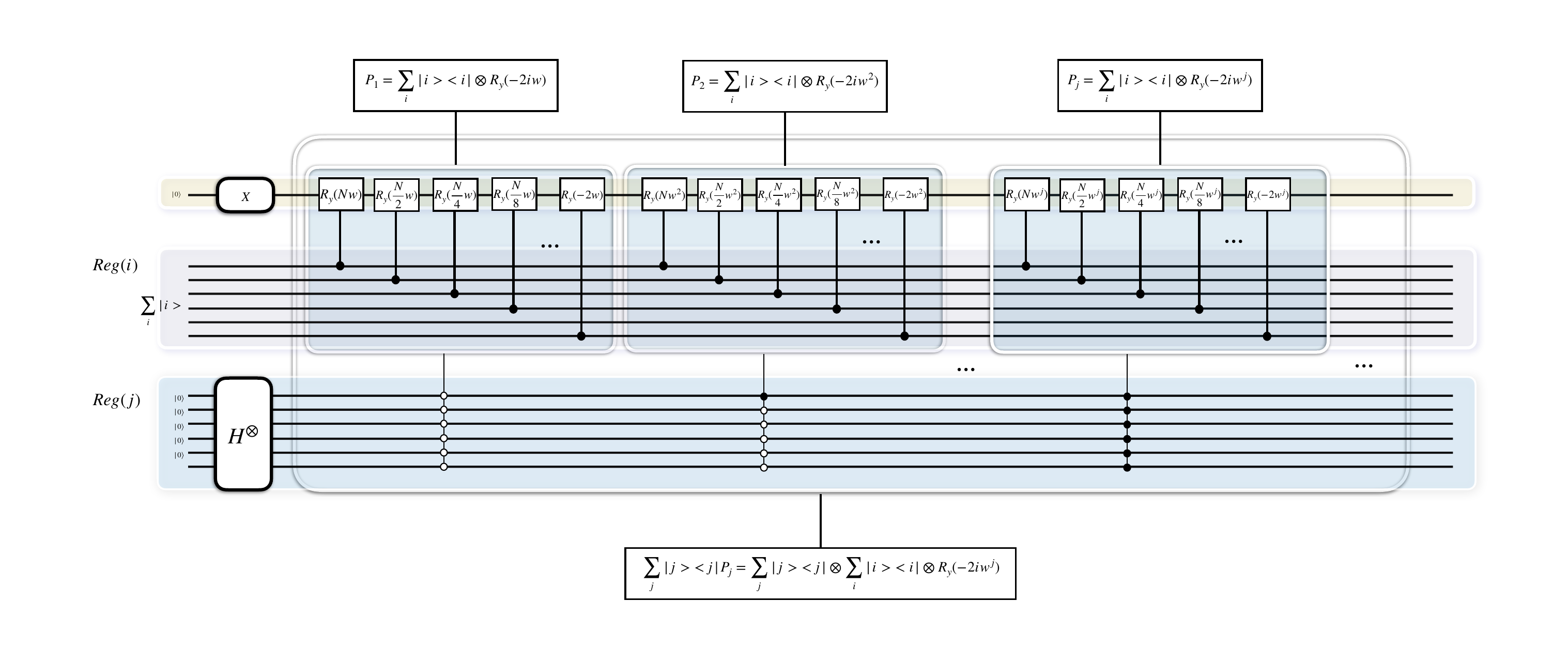}
   \caption{\textit{Quantum circuit for positional encoding}  Build upon the two registers $Reg(i)$ and $Reg(k)$ hosting the index $i$ and $k$ respectively (illustrated in Fig.~\ref{qattention1}), we set up a register $Reg(j)$ hosting the index $j$ below $Reg(i)$ and an ancillary qubit above $Reg(i)$. The blue boxes that group the series of controlled $R_Y$ gates implement the following unitaries: $P_j = \sum_i \ket{i}\bra{i} \otimes R_y(-2iw^j)$, each of which is controlled by the qubits in $Reg(j)$ and the entire controlled sequences grouped in the transparent box implement the unitary $U_c = \sum_j \ket{j}\bra{j} \otimes P_j = \sum_j \ket{j}\bra{j} \otimes \sum_i \ket{i}\bra{i} \otimes R_y(-2iw^j)$. The whole circuit implements $U_{\boldsymbol{P}}$. Note that in this figure, $N=-n$.}
    \label{position}
\end{figure}

Denote the unitary that prepares $\left|\psi_{\boldsymbol{P}}\right\rangle$ from $\sum_{i=1}^{n} |i\rangle\ket{0}$ as $U_{\boldsymbol{P}}$, then $\left|\psi_{\boldsymbol{X'}}\right\rangle$ can be achieved by applying LCU to $U_{\boldsymbol{X}}$,$U_{\boldsymbol{P}}$. As the construction of $U_{\boldsymbol{X}}$ is given by the CQSP mentioned in Section \ref{input}, we can focus on the construction of $U_{\boldsymbol{P}}$. Next, we present the construction of $U_{\boldsymbol{P}}$ as depicted in Fig.\ref{position}.

Build upon the two registers $Reg(i)$ and $Reg(k)$ hosting the index $i$ and $k$ respectively (illustrated in Fig.~\ref{qattention1}), in Fig.~\ref{position}, we set up a register $Reg(j)$ hosting the index $j$ below $Reg(i)$ and an ancillary qubit above $Reg(i)$. For reasons that will be clear soon, we combine $Reg(j)$ and the ancillary qubit as a single register which coincides with $Reg(k)$. The blue boxes that group the series of controlled $R_Y$ gates implement the following unitaries:

\begin{equation}
P_j = \sum_i \ket{i}\bra{i} \otimes R_y(-2iw^j)
\end{equation}

each of which is controlled by the qubits in $Reg(j)$, and the entire controlled sequences grouped in the transparent box implement the unitary 

\begin{equation}
U_c = \sum_j \ket{j}\bra{j} \otimes P_j = \sum_j \ket{j}\bra{j} \otimes \sum_i \ket{i}\bra{i} \otimes R_y(-2iw^j)
\end{equation}

The Hadamard gates and $X$ gate before $U_c$ transform the input state to the state $\sum_j \ket{j} \otimes \sum_i \ket{i} \otimes \ket{1}$, after $U_c$, it becomes

\begin{align}
U_c (\sum_j \ket{j} \otimes \sum_i \ket{i} \otimes \ket{1})= \sum_j \ket{j} \otimes \sum_i \ket{i} \otimes R_y(-2iw^j)\ket{1} 
\end{align}

By placing $Reg(j)$ and the ancillary qubit next to each other we can rewrite the above state as:

\begin{align}
 \sum_i \ket{i} \otimes \sum_j \ket{j} \otimes R_y(-2iw^j)\ket{1} 
\\= \sum_i \ket{i} \otimes \sum_j \ket{j} \otimes (\sin(iw^j)\ket{0} + \cos(iw^j)\ket{1}).
\end{align}

Combining $Reg(j)$ and the ancillary qubit as a single register that coincides with $Reg(k)$, the computational basis transform as $\ket{j}\ket{0} \to \ket{2j}$,$\ket{j}\ket{1} \to \ket{2j+1}$  and we can write the output state from the circuit in Fig.\ref{position} as:

\begin{equation}
\ket{\text{output}}		= \sum_i \ket{i} \otimes \sum_j (\sin(iw^j)\ket{2j} + \cos(iw^j)\ket{2j+1} )
\label{outlet}
\end{equation}
set $w=\frac{1}{10000^{\frac{1}{d}}}$, and from Eqn.~\ref{definition},\ref{psip},\ref{pe1},\ref{outlet} we have

\begin{equation}
\ket{\text{output}}		= \left|\psi_{ \boldsymbol{P}}\right\rangle
\end{equation}

That is, the circuit in Fig.\ref{position} implements $U_{\boldsymbol{P}}$.
\section{Masked-Attention}\label{mask}
This section presents the masking operation in attention and its quantum implementation.\newline

The masked attention is defined as:\cite{ghojogh2020attention}

$$
\begin{aligned}
\mathbb{R}^{r \times n} \ni \boldsymbol{Z}_{m} & :=\operatorname{maskedAttention}(\boldsymbol{Q}, \boldsymbol{K}, \boldsymbol{V}) \\
& =\boldsymbol{V} \operatorname{softmax}\left(\frac{1}{\sqrt{p}}\left(\boldsymbol{Q}^{\top} \boldsymbol{K}+\boldsymbol{M}\right)\right),
\end{aligned}
$$

where the mask matrix $\boldsymbol{M} \in \mathbb{R}^{n \times n}$ is:

$$
\boldsymbol{M}_{ij}:= \begin{cases}0 & \text { if } j \leq i \\ -\infty & \text { if } j>i\end{cases}
$$

For positions \(j \leq i\) in a sequence (representing current or previous words), the mask doesn't alter the softmax output, this allows these positions to contribute to the softmax computation. In contrast, for positions \(j > i\) (corresponding to future words in the sequence),  the nature of softmax function indicates that the mask effectively nullifies their contribution as \(e^{-\infty}\) is 0. This selective masking ensures that the model's attention is appropriately focused on the relevant parts of the sequence: considering past and present words while ignoring future words.\newline

Denoting  $\boldsymbol{A}^{\textsf{Mask}}_0\equiv \operatorname{softmax}\left(\frac{1}{\sqrt{p}} \left(\boldsymbol{Q}^{\top} \boldsymbol{K}+\boldsymbol{M}\right)\right)$, we have its elements as

$$
{\boldsymbol{A}^{\textsf{Mask}}_0}_{ij}:= \begin{cases}{\boldsymbol{A}_0}_{ij} & \text { if } j \leq i \\ 0 & \text { if } j>i\end{cases}
$$
where ${\boldsymbol{A}_0}_{ij}$ is the elements of $\boldsymbol{A}_0\equiv \operatorname{softmax}\left(\frac{1}{\sqrt{p}} \boldsymbol{Q}^{\top} \boldsymbol{K}\right)$.\newline

In the quantum case, we aim to implement an alternative version of  $\boldsymbol{A}^{\textsf{Mask}}_0$ denoted as  $\boldsymbol{A}^{\textsf{Mask}}$ whose elements are defined as

$$
{\boldsymbol{A}^{\textsf{Mask}}}_{ij}:= \begin{cases}{\boldsymbol{A}}_{ij} & \text { if } j \leq i \\ 0 & \text { if } j>i\end{cases}
$$

where ${\boldsymbol{A}}_{ij}$ is the elements of $\boldsymbol{A}\equiv \boldsymbol{Q}^{\top} \boldsymbol{K}$.\newline

For masked-attention, we aim to design a quantum circuit implementing the following computation:

\begin{equation}
 	 \boldsymbol{Z'}_m =\boldsymbol{W}_{V}^{\top} \boldsymbol{X} \boldsymbol{A}^{\textsf{Mask}},
\label{mattentionq}	   
\end{equation}

This can be done in a similar way as in the case of self-attention where we implemented Eqn.\ref{attentionq}, the only difference is that we now need the block-encoding of ${\boldsymbol{A}^{\textsf{Mask}}}^{\top}$ (instead of $\boldsymbol{A}^{\top}$) whose elements are

$$
\Lambda_{ij}^{\textsf{Mask}}:={\boldsymbol{A}^{\textsf{Mask}}}^{\top}_{ij}:= \begin{cases} 0& \text { if } j < i \\ {\boldsymbol{A}}_{ji}  & \text { if } j \geq i\end{cases}
$$

Similar to the case of self-attention, let \( \hat{\Lambda}^{\textsf{Mask}} \geq \max_{i,j} |\Lambda^{\textsf{Mask}}_{ij}| \) and \( \tilde{\Lambda}^{\textsf{Mask}}_{ij} \) is defined to be the (exact) b-qubit description of \( \Lambda^{\textsf{Mask}}_{ij} / \hat{\Lambda}^{\textsf{Mask}} \). The block-encoding of ${\boldsymbol{A}^{\textsf{Mask}}}^{\top}$ can be constructed using lemma 3.2 from Ref.\cite{nguyen2022block}, given the following oracle

\begin{equation}
	O_{{\boldsymbol{A}^{\textsf{Mask}}}^{\top}} : |i\rangle |j\rangle |0\rangle^{\otimes b} \rightarrow |i\rangle |j\rangle |\tilde{\Lambda}^{\textsf{Mask}}_{ij}\rangle,
	\end{equation}
where \( 0 \leq i,j < n \). \newline

This oracle $O_{{\boldsymbol{A}^{\textsf{Mask}}}^{\top}} $ can be constructed by conditionally applying $O_{\boldsymbol{A}^{\top}}$: on the registers hosting $|i\rangle |j\rangle$, set up a circuit comparing the values of $i,j$ with the result stored in an extra ancillary qubit (using Claim 3.1 in Ref.\cite{landman2021quantum}). Then using this ancillary qubit as control qubit,  apply controlled $O_{\boldsymbol{A}^{\top}}$ if $j \geq i$.

\section{Block-encoding}\label{blocke}

\begin{figure}[h!]
    \centering
    \includegraphics[width=\linewidth]{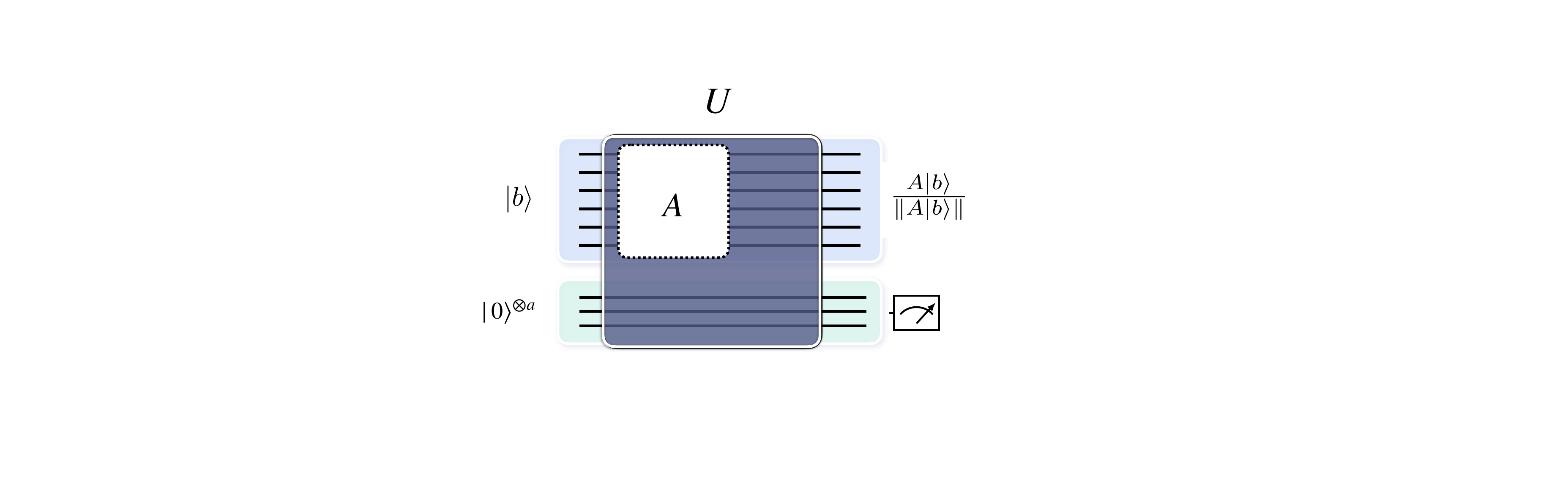}
   \caption{\textit{Block-encoding } \(U\), the Block-encoding of a matrix A, can be considered as a probabilistic implementation of \(A\): applying the unitary \(U\) to a given input state \(|0\rangle^{\otimes a}|b\rangle\), measuring the first \(a\)-qubit register and post-selecting on the outcome \(|0\rangle^{\otimes a}\), we get state proportional to \( A|b\rangle\) in the second register.}
    \label{Block}
\end{figure}

Block encoding is a powerful modern quantum algorithmic technique that is employed in a variety of quantum algorithms for solving linear algebra problems on a quantum computer\cite{sunderhauf2024block}.  A unitary $U$ is a block encoding of a not-necessarily-unitary square matrix $A$ ($A$ is scaled to satisfy $\|A\|_2\leq 1$\cite{sunderhauf2024block}) if $A$ is encoded in the top-left block of the unitary $U$ as:
$$
U=\left[\begin{array}{cc}
A & . \\
\cdot & \cdot
\end{array}\right],
$$

where the $\cdot$ symbol stands for a matrix block. Equivalently, we can write 

\begin{equation}
    A=\left(\left\langle 0\right|^{\otimes a} \otimes I\right) U\left(|0\rangle^{\otimes a} \otimes I\right) 
\end{equation}

where $a$ is the number of ancilla qubits used for the block encoding of $A$. $U$ can be considered as a probabilistic implementation of $A$: by applying the unitary $U$ to an input state $ |0\rangle^{\otimes a} |b\rangle$, measuring the first $a$-qubit register and post-selecting on the outcome $ |0\rangle^{\otimes a}$, we obtain a state that is proportional to $A |b\rangle$ in the second register. This can be illustrated in Fig.\ref{Block}. \newline

The circuit implementation of Block-encoding in general can be constructed using the Linear Combination of Unitaries(LCU)\cite{childs2012hamiltonian} technique.\cite{lin2022lecture}

\section{Parametrized quantum circuit for implementing trainable linear layer}\label{para}

Classical neural networks are fundamentally built on the structure of multilayer perceptrons which involve layers of trainable linear transformations and element-wise nonlinear transformations (activation functions such as ReLU, sigmoid, or tanh).  On the other hand, Quantum Neural Networks (QNNs), which are often defined as parametrized quantum circuits with a predefined circuit ansatz, do not naturally exhibit this kind of structure. In QML literature, a QNN, denoted as $U(\thv)$, often has a structure of layers $L$ of the form \cite{larocca2021theory}
\begin{equation}\label{eq:PSA_ansatz}
    U(\thv)=\prod_{l=1}^LU_l(\thv_l)\,, \quad U_l(\thv_l)=\prod_{k=1}^K e^{-i \theta_{lk}H_k}\,,    
\end{equation}
where the index $l$ represents the layer, and the index $k$ covers the Hermitian operators $H_k$ that generates the unitaries in the circuit ansatz, $\thv_{l}=(\theta_{l1},\ldots\theta_{lK})$ represents the parameters in a single layer, and $\thv=\{\thv_{1},\ldots,\thv_L\}$ represents the collection of adjustable parameters in the QNN. Examples of circuit ansatz represented by Eq.~\ref{eq:PSA_ansatz} include: the hardware-efficient ansatz~\cite{kandala2017hardware}, quantum alternating operator ansatz~\cite{hadfield2019quantum}, and quantum optimal control Ansatz~\cite{choquette2020quantum}, among others. \newline

 In machine learning, a linear layer~\cite{goodfellow2016deep} is a fundamental component of neural networks architectures that maps input vectors to output vectors through affine transformation: Given an input vector $\mathbf{x} \in \mathbb{R}^n$, the linear layer transforms it to an output vector $\mathbf{y}\in \mathbb{R}^m$ using a weight matrix \(W \in \mathbb{R}^{m \times n}\) and an optional bias vector \(\mathbf{b} \in \mathbb{R}^m\) as \(\mathbf{y} = W\mathbf{x} + \mathbf{b}\), and \(W\), \(b\) contain the trainable parameters. \newline

 The input vector $\mathbf{x}$ can be encoded in the amplitudes of a $s$-qubit quantum state ($n=2^{s}$) $\left|\psi_{x}\right\rangle$ as
$
\left|\psi_{x}\right\rangle=\sum_{i=1}^{n} x_{i}|i\rangle
$
where $ x_{i}$ is the $i$-th element of $\mathbf{x}$(after normalization), and $|i\rangle$ is the $i$-th computational basis state. Applying a parameterized quantum circuit $U(\thv) \in \mathbb{C}^{n \times n}$ on $\left|\psi_{x}\right\rangle$ be interpreted as a special type of linear layer (represented as a square matrix) on $\mathbf{x}$. In this paper, we utilize parameterized quantum circuits as in Eq.~\ref{eq:PSA_ansatz} for implementing such linear layers. Note that in case where the weight matrix \(W \in \mathbb{R}^{m \times n}\) is rectangular ($m \neq n$), by default we adjust the dimension of the output vector to be the same as the input vector in our quantum architecture, without specifying the adjustment in the prior description of the classical architecture.\newline

It should be emphasized that the trainable circuit parameters $\thv$ are not equivalent to the weights in the weight matrix, but rather are parameterized by them, as in $W (\thv)$. Ref \cite{larocca2021theory} contains a discussion of the parametrization.

\end{document}